\documentclass{ametsocV6.1}
\usepackage{graphicx}
\usepackage{subcaption}
\graphicspath{{figures/}}
\usepackage{longtable}
\usepackage{float}
\usepackage{tikz}
\usepackage{lineno}

\usepackage{nicefrac, xfrac}
\usepackage{xcolor}
\title{High-resolution vertical wind and turbulence measurements with quadcopter uncrewed aerial systems: wind tunnel calibration and field validation \\}

\nolinenumbers
\authors{Johannes Kistner\aff{a,b}\correspondingauthor{Johannes Kistner, johannes.kistner@dlr.de \\ \textcolor{red}{This work has been submitted to Journal of Atmospheric and Oceanic Technology. Copyright in this work may be transferred without further notice.}}, Julian Jüchter\aff{b,c}, Norman Wildmann\aff{a}  }

\nolinenumbers
\affiliation{\aff{a}{Deutsches Zentrum für Luft- und Raumfahrt, Institut für Physik der Atmosphäre, Oberpfaffenhofen, Germany}\\ \aff{b}{Carl von Ossietzky Universität Oldenburg, School of Mathematics and Science, Institute of Physics} \\ \aff{c}{ForWind - Center for Wind Energy Research, Küpkersweg 70, 26129 Oldenburg, Germany} \nolinenumbers}

\nolinenumbers
\abstract{
The SWUF-3D drone fleet is used in the atmospheric boundary layer (ABL) for in situ turbulence measurements of complex flows, such as in mountainous terrain or wind turbine wakes. 
Previous calibrations for measuring vertical wind speed $w$ using the drones' avionics data were performed on field data, limiting applicability to low winds ($\leq 8~\mathrm{m}\,\mathrm{s}^{-1}$) and being prone to high uncertainties.
To overcome these limitations, we calibrate $w$ measurement in a wind tunnel and validate it in field measurements. 
Calibration is performed in a wind tunnel with an active grid used to deflect horizontal flow into the vertical. This wind is measured with a multi hole probe, while wind forces acting on the drone are determined from the avionics data, allowing an empirical relationship between these quantities. 
For validation, we conduct comparative fleet measurements with up to 10 drones simultaneously around an array of meteorological masts equipped with sonic anemometers.
The results show high accuracy for turbulence statistics: the variance determination for $w$ has a root mean square error (RMSE) of 0.12~$\mathrm{m^2\,s^{-2}}$ and a normalized RMSE (nRMSE) of 17.0~\%, for the horizontal wind components the RMSEs are $\sim$0.3~$\mathrm{m^2\,s^{-2}}$ and nRMSEs $\sim$25~\%. The RMSEs for the covariances of the components are $<\,$0.3~$\mathrm{m^2\,s^{-2}}$. The variance spectra of $w$ measured with drones and reference sensors agree in all frequency ranges, the RMSE for covariances between different measurement points is $\sim$0.1~$\mathrm{m^2\,s^{-2}}$.
Accurate $w$ retrieval at all wind speeds sustainable by the drone enables studies of strongly three-dimensional flows, supports eddy-covariance flux estimation, enables resolving diurnal turbulence evolution in the ABL, and improves spatial turbulence characterization.
\nolinenumbers}
\nolinenumbers
\begin{document}
\nolinenumbers
\maketitle
\nolinenumbers
%
%
%
\nolinenumbers
\statement
We use a fleet of multicopter drones to measure wind. This data is used to improve models of the atmosphere in complex conditions, e. g. in mountainous terrain or around wind turbines. To improve the accuracy and capabilities of this wind measurement, we conducted unique experiments with the drones in a wind tunnel under various conditions. For evaluation of this improvement, we conducted field measurements in flights with the drone fleet in comparison to reference sensors. In the presented work, we were able to extend our wind measurement capabilities and show how accurately it works in different situations. This enables better meteorological measurements to provide data, and a more detailed analysis of this measurement data.

\section{Introduction}
\nolinenumbers
Measuring wind and turbulence in the atmospheric boundary layer (ABL) is important for various applications in science and industry, e.g., in meteorology and wind energy. Uncrewed aerial systems (UASs), also known as drones, are now commonly used for  in situ measurements, as they are more flexible, less dependent on ground conditions, and can be used at higher altitudes than measurement infrastructure such as masts.
The UASs used can be divided into fixed-wing drones \citep{Wildmann2015, Rautenberg2019, Bramati2025} and copter drones. The latter are characterized by vertical takeoff and landing capabilities, high flexibility of movement in all spatial directions, and the ability to hover. Measurements with copter UASs can again be subdivided into the direct measurement method, with the drone as a carrier platform for anemometers \citep{Thielicke2021, Ghirardelli2025, Yang2025}, and the indirect measurement method with the drone itself as a wind sensor, using avionics data \citep{Neumann2015, Palomaki2017, Brosy2017, Segales2020, GonzalezRocha2020}.
\newline
The SWUF-3D drone fleet \citep{Wetz2021} is using the latter method. We have operated this fleet so far with up to 30 UASs hovering simultaneously to measure turbulence in the ABL in a spatially distributed pattern. The indirect method is advantageous given this high number of drones as it does not require an equally high number of additional anemometers. In order to be able to measure highly three-dimensional processes with the fleet such as mountain boundary layer phenomena \citep{Pfister2024a} and wind turbine wakes \citep{Wildmann2024, Wildmann2025}, and to enable flux estimation \citep{Wildmann2025a}, it is necessary to measure the full 3D wind vector, especially the vertical wind component.
\newline
First studies on measuring the vertical wind component using the indirect method have already been carried out:  
\cite{Segales2025} use a linear extended state observer of the aerodynamic drag as an estimate of the vertical wind vector. \cite{Zhu2025} present a machine learning approach on avionics data in which the 3D wind vector is determined and vertical winds $<$1.5$~\mathrm{m~s^{-1}}$ were measured. \cite{Wildmann2022} have demonstrated the first method that enables vertical wind and turbulent flux measurement with multicopter UAS without external sensors. They performed calibrations in field measurements for the SWUF-3D fleet. However, this method is subject to certain limitations: on the one hand, it can only be used for measurements at horizontal wind speeds of up to 8$~\mathrm{m~s^{-1}}$, and on the other hand, due to the high distance between UAS and reference sensor in the field calibration the measurement is prone to high uncertainties. To overcome these limitations, we perform calibrations in a wind tunnel and validate these in field measurements.
\newline
We proceed in a similar manner to \cite{Kistner2024}, where the focus was limited to determining horizontal wind speeds using accelerations. For the calibrations for vertical wind measurements, the determination of forces caused by the propellers must also be taken into account in the calibrations.
\newline
\cite{Neumann2015} and \cite{Hattenberger2022} also calibrated indoors or in a wind tunnel, but for horizontal wind only. \cite{GonzalezRocha2019} also carried out a characterization for individual rotors in a wind tunnel, but did not perform any calibrations for wind measurement with the entire drone. \cite{Marino2015} calibrated their measurement method for the power consumption of the rotors in a wind tunnel, but they conclude that its practicality is limited. \cite{Thielicke2021} performed a wind tunnel calibration of their system, which consists of an sonic anemometer on a multicopter, which therefore is a case of the direct wind measurement technique. 
\newline
Thus, this work is the first time that vertical wind measurement calibration has been performed in a wind tunnel with a UAS as complete system in flight.
Our hypothesis is that wind tunnel calibration will enable us to overcome the above-mentioned limitations and present the first deterministic measurement method for the full 3D wind vector using a multicopter UAS for all its operating conditions.  
We validate this by conducting field measurements with the fleet in comparison to measurement masts. This hypothesis leads to the following research questions:

\begin{enumerate}
    \item How feasible, stable and reproducible is a wind tunnel calibration for measuring the vertical wind component?
    \item In what extent can the wind tunnel calibration be transferred to fleet measurements?
    \item What are the absolute and relative uncertainties of the calibrated 3D wind measurements, especially of the vertical wind component, under typical operating conditions?
    \item To what extent can the wind tunnel calibration improve vertical wind measurement compared to the previous field calibration?   
\end{enumerate}

This work aims to improve the calibration of 3D wind measurement and is intended to validate the measurement method in the field to give conclusions about measurement capabilities when measuring with the entire fleet. 
\newline
The next chapter describes the multicopter UASs used, the wind tunnel with active grid and reference sensors, and the experimental setup in the research wind farm. The third chapter describes the methods used to determine turbulence parameters and to quantify uncertainties when measuring these, and to conduct the calibrations and validations. The results are then presented in chapter 4 and discussed in chapters 5 and 6. 

 \section{System Description and Experimental Setup}

 \subsection{UAS System}

\begin{figure}[H]
  \centering
  \includegraphics[width=0.48\textwidth]{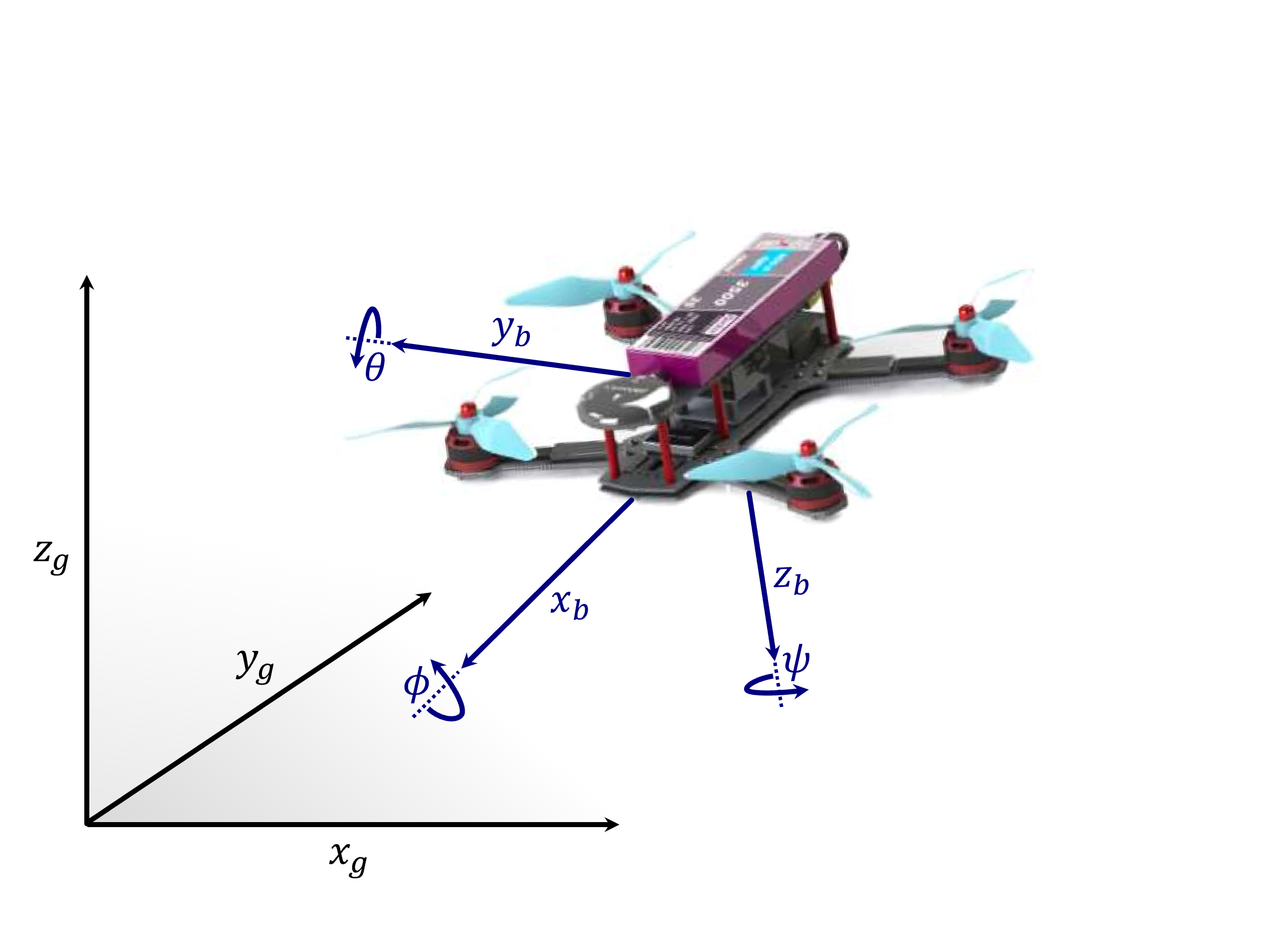} 
  \caption{The geodetic coordinate system (black) and the body-fixed coordinate system with Euler angles (blue) of the QAV250 drone in the 5.5$^{\prime\prime}$ setup (adapted from \citeauthor{Hofmann2025}, \citeyear{Hofmann2025}).}
  \label{fig:coordinate_system}
\end{figure}

 Our drone is based on the Holybro QAV250 quadrotor airframe with the Pixhawk 4 Mini autopilot, on which we use a custom version of PX4 v14.1.0. The drone measures 0.25~m from motor to motor and weighs 0.655~kg including the battery. This enables a flight time of approximately 15~minutes with standard 5.0$^{\prime\prime}$ propellers. To extend the flight time to approximately 20~minutes, we have installed 5.5$^{\prime\prime}$ propellers. 
 \newline
 We perform our measurements in hover flight with the drone operating in weather vane mode, i. e. it always yaws the drone towards the main wind. Positioning is ensured in the field via global navigation satellite system (GNSS) and barometer. In the wind tunnel, the drone maintains its horizontal position via an optical flow sensor and its altitude via a rangefinder, both of which are installed on the TF-01 module by Robofusion.
 \newline
 Our objective is to determine the wind in the geodetic coordinate system $\textbf{\textit{V}}_{\mathrm{g}}$. The avionics data we use to determine the wind is only available in the drone's body-fixed coordinate system (see Fig \ref{fig:coordinate_system}). Therefore, we first determine the wind speed decomposed into its 3D components in the body-fixed coordinate system (i. e.  $u_{\mathrm{b}}$,  $v_{\mathrm{b}}$,  and $w_{\mathrm{b}}$), transform it into the geodetic coordinate system (via \textbf{\textit{R}}) and correct it for movements of the UAS ($ u_{\mathrm{loc}}$, $ v_{\mathrm{loc}}$, and $w_{\mathrm{loc}}$) measured by the localization system (i. e. GNSS or optical flow and rangefinder): 

\begin{equation}\label{eq:gesamt}
\textbf{\textit{V}}_{\mathrm{g}} = \begin{pmatrix} u_{\mathrm{g}} \\ v_{\mathrm{g}} \\ w_{\mathrm{g}} \end{pmatrix}
=  \textbf{\textit{R}}(\phi,\theta,\psi)
\begin{pmatrix}
 u_{\mathrm{b}}\\
 v_{\mathrm{b}}\\
 w_{\mathrm{b}}
\end{pmatrix}
+
\begin{pmatrix}
 u_{\mathrm{loc}}\\
 v_{\mathrm{loc}}\\
 w_{\mathrm{loc}}
\end{pmatrix}
\end{equation}
\begin{equation} \label{eq:rot_matrix}
    \textbf{\textit{R}}(\phi,\theta,\psi)=
\begin{bmatrix}
 \cos\theta\cos\psi & \cos\psi\sin\theta\sin\phi - \cos\phi\sin\psi & \cos\psi\sin\theta\cos\phi + \sin\phi\sin\psi \\[0.75em]
 \cos\theta\sin\psi & \cos\theta\cos\psi + \sin\theta\sin\phi\sin\psi & -\sin\phi\cos\psi + \sin\theta\cos\phi\sin\psi \\[0.75em]
 \sin\theta & -\cos\theta\sin\phi & -\cos\theta\cos\phi
\end{bmatrix}
\end{equation}
Eq. \ref{eq:rot_matrix} inserted into \ref{eq:gesamt} gives:
\begin{align} 
 u_{\mathrm{g}} &= \cos\theta\cos\psi\,u_{\mathrm{b}}
 + (\cos\psi\sin\theta\sin\phi - \cos\phi\sin\psi)\,v_{\mathrm{b}} \notag\\
 &\quad + (\cos\psi\sin\theta\cos\phi + \sin\phi\sin\psi)\,w_{\mathrm{b}}+u_{\mathrm{loc}}, \label{eq:1.3a} \\[0.5em]
 v_{\mathrm{g}} &= \cos\theta\sin\psi\,u_{\mathrm{b}}
 + (\cos\theta\cos\psi + \sin\theta\sin\phi\sin\psi)\,v_{\mathrm{b}} \notag\\
 &\quad + (-\sin\phi\cos\psi + \sin\theta\cos\phi\sin\psi)\,w_{\mathrm{b}}+v_{\mathrm{loc}}, \label{eq:1.3b} \\[0.5em]
 w_{\mathrm{g}} &= \sin\theta\,u_{\mathrm{b}}
 - \cos\theta\sin\phi\,v_{\mathrm{b}}
 - \cos\theta\cos\phi\,w_{\mathrm{b}}+w_{\mathrm{loc}} ~. \label{eq:1.3c}
\end{align}

 \subsection{Wind Tunnel}
 
\begin{figure}[H]
  \centering
  \includegraphics[width=0.48\textwidth]{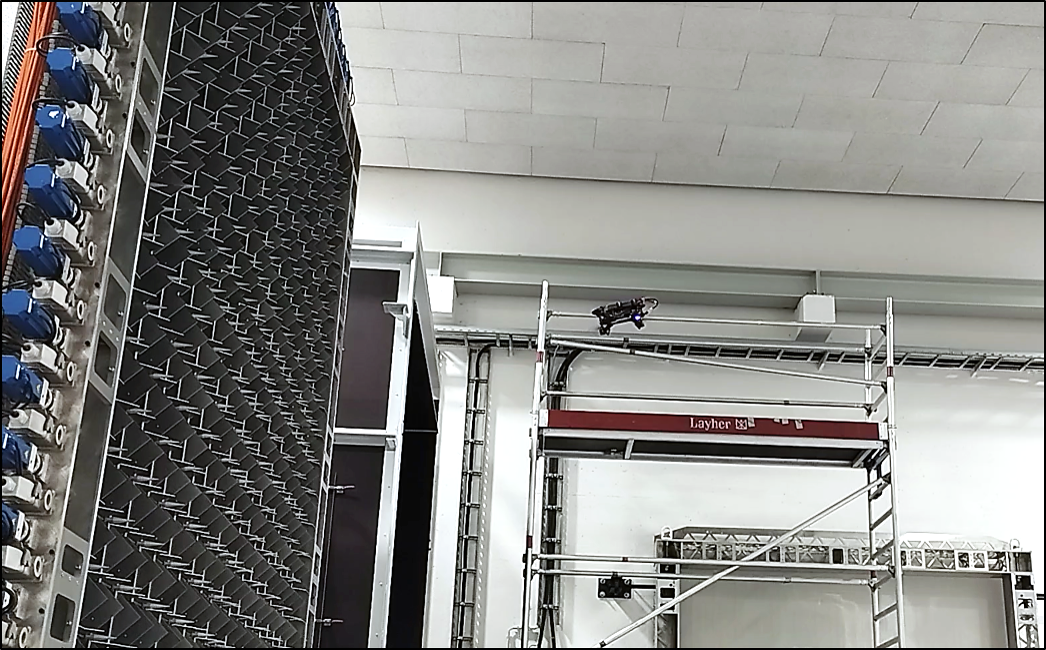} 
  \caption{QAV250 drone in the 5.0$^{\prime\prime}$ setup hovering in front of the wind tunnel nozzle equipped with the active grid.}
  \label{fig:hover_vertical}
\end{figure}

The Oldenburg wind tunnel features a 3$\times$3~m$^2$ nozzle and a 30~m long test section. At the nozzle, an active grid with 80 individually controllable shafts can be installed \citep{Neuhaus2021}. Attached to these shafts are 880 flaps, each measuring 0.14 m, which can be rotated to modify the local blockage of the wind tunnel. Using the active grid, the overall blockage ratio of the wind tunnel can be adjusted between 21\% and 92\%. The grid consists of 40 horizontal and 40 vertical shafts, each extending from the outside to the center of the active grid. Defined control protocols allow the shaft motions to be reproduced precisely during experiments, enabling a high level of repeatability that is not achievable in field measurements. Both turbulence intensity and flow direction can be influenced by the active grid.
\newline
The wind tunnel can be operated in either closed or open configuration. For the calibration flights conducted here, the open test section is used, as it provides more maneuvering space for the drones and enables the active grid to generate stronger vertical winds. In the open test section, air exits the nozzle at speeds of up to 32~$\mathrm{m}~\mathrm{s}^{-1}$ into a large laboratory hall without direct lateral confinement at the sides of the nozzle. In both closed and open configurations, environmental conditions such as temperature, humidity, and pressure can be kept stable.
\newline
During the experiments, the drone was positioned in the center of the nozzle and at a short distance downstream of the active grid. This ensures that the boundary layer between moving and stationary air at the sides of the flow does not have a significant influence on the measurements.
\newline
For horizontal wind calibration, we use a Prandtl probe as reference, which measures simultaneously with the UAS in the airflow. For the vertical wind component, we use a multi hole probe, which we place at the same location as the drone's hovering position (see Fig.~\ref{fig:hover_vertical}) and measure the same time series in different runs as the UAS. This is only possible because of the high accuracy of the wind tunnel, repeating the same pre-defined flow pattern.

 \subsection{Wind Farm}
\begin{figure}[H]
  \centering
  \includegraphics[width=0.48\textwidth]{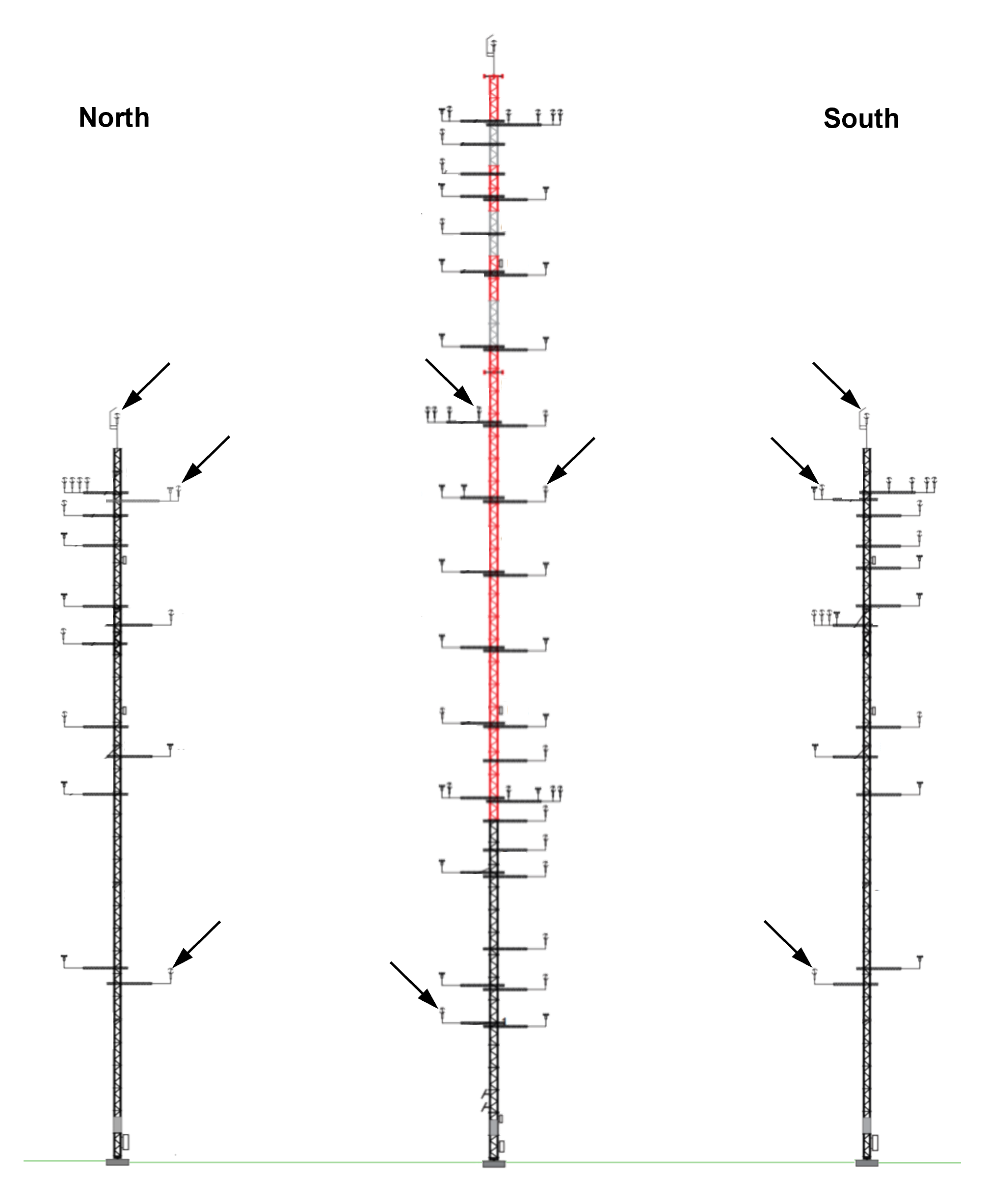} 
  \caption{Schematic representation of the measurement mast array. The arrows indicate the sonic anemometers selected as references at heights of 20, 25, 90, and 100~m.}
  \label{fig:metmast_sensors}
\end{figure}

We use the sensors of the measurement mast array at the DLR WiValdi research wind park as a reference for validation of our field measurements. The array of masts at the WiValdi research park is a unique facility that was build to measure the spatial representation of turbulence on scales that are most relevant for wind turbines. The UAS fleet is an excellent extension to this facility, as it can extend the in situ measurements to another dimension and extent it further in horizontal and vertical direction. In this study, we want to validate that the drones are capable of providing a comparable quality of spatial turbulence to the mast on a wide range of scales. The array consists of three measurement masts spaced 50~m apart, which are aligned along the geographical $10^{\circ}$-$190^{\circ}$ axis, and of which the middle mast is 150~m in height and the outer masts are 100~m in height. Cup anemometers and 3D sonic anemometers are installed at heights of 20~m up to the tops of the measurement masts, of which we use the following sonic anemometers as a reference:
On the outer masts, the sensors used are at 25~m, 90~m, and 100~m; on the middle mast, they are at 20~m, 90~m, and 150~m. The selection of the respective reference sensors is shown in Fig.~\ref{fig:metmast_sensors}.

\begin{figure}[H]
  \centering
  \begin{subfigure}{0.48\textwidth}
    
    \includegraphics[width=\linewidth]{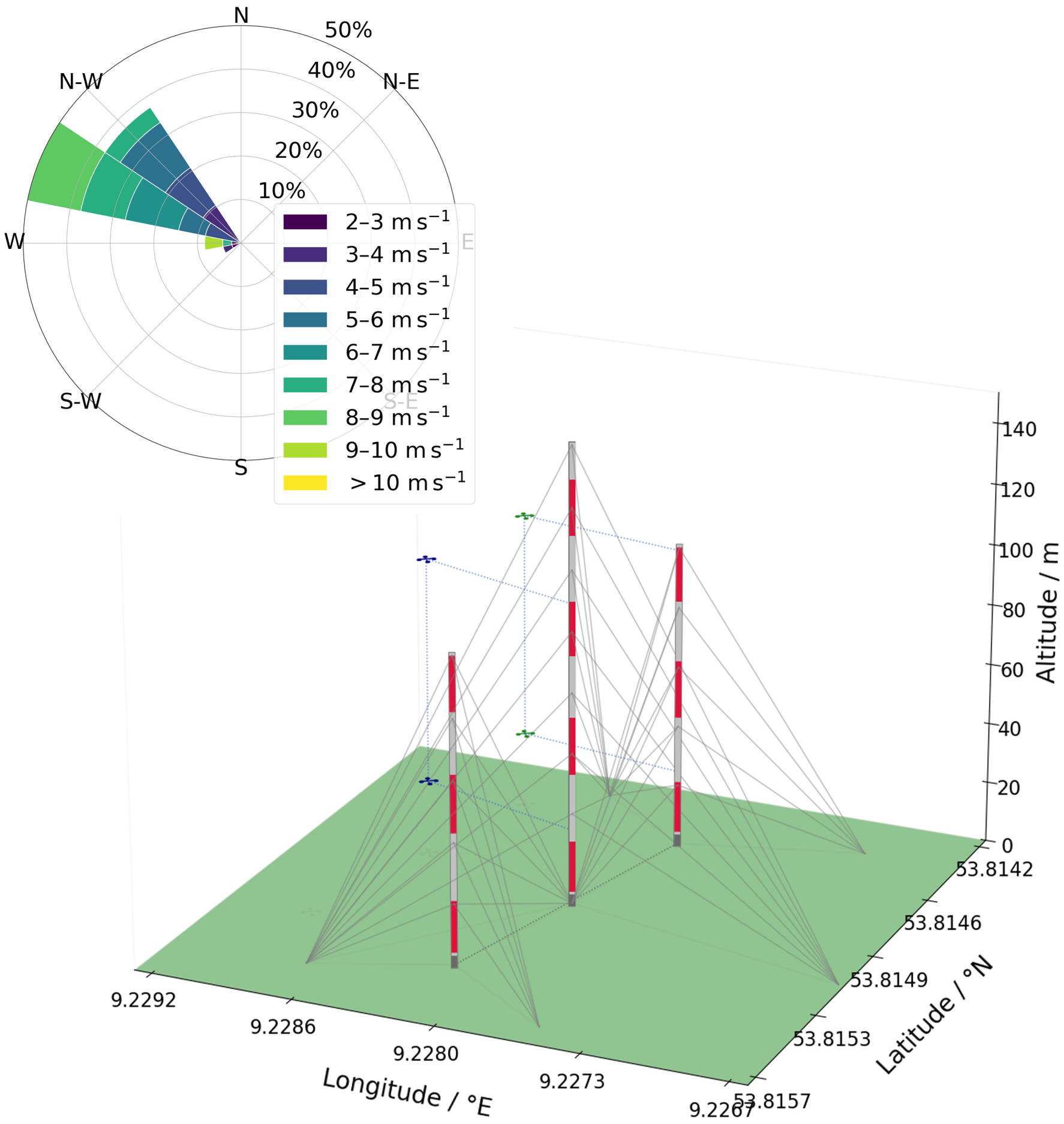}
    \caption{Flight pattern for setup comparison}
    \label{fig:patterns_windroses_a}
  \end{subfigure}
  \hfill%
  \begin{subfigure}{0.48\textwidth}
    
    \includegraphics[width=\linewidth]{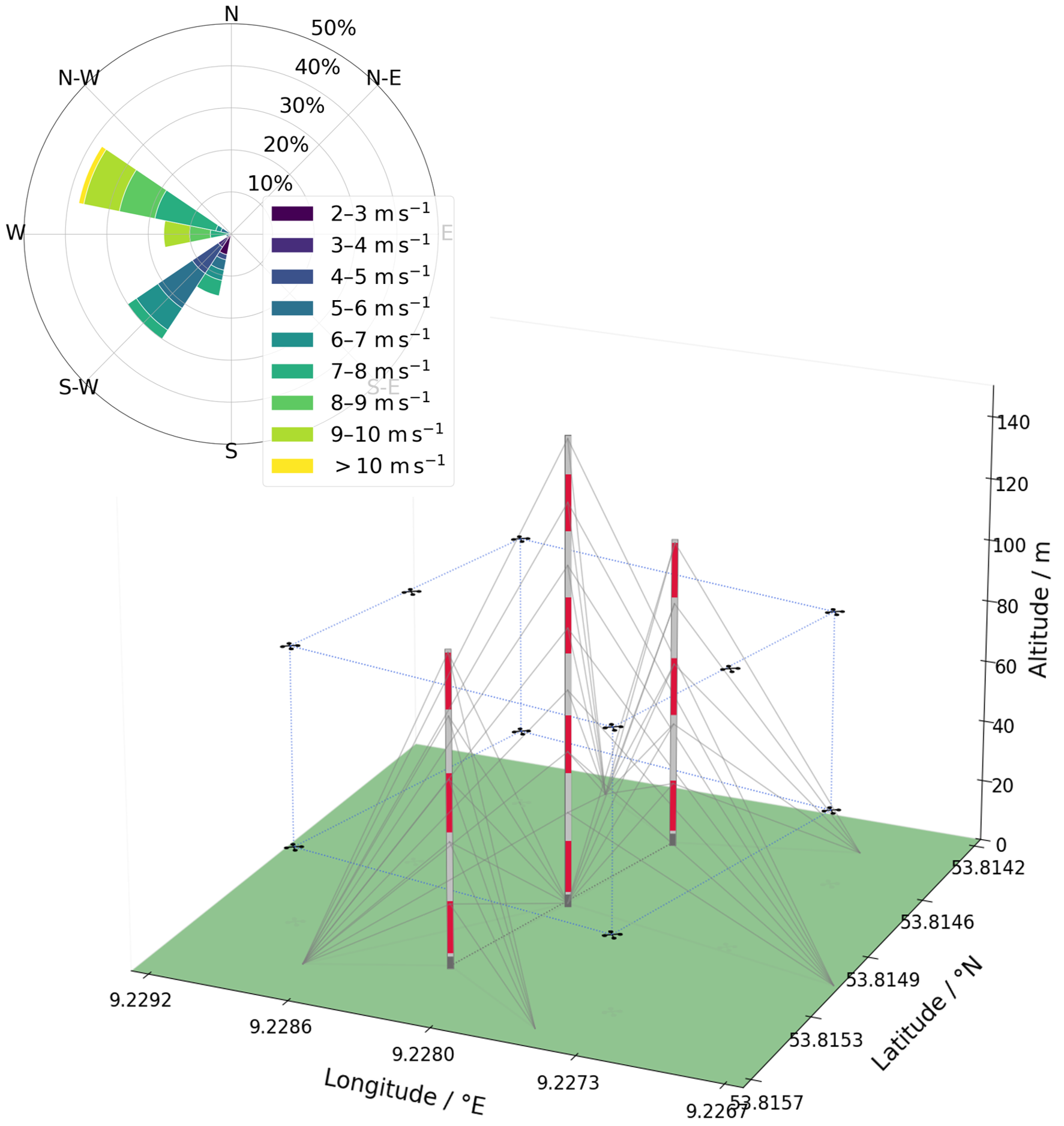}
    \caption{Box-shaped flight pattern}
    \label{fig:patterns_windroses_b}
  \end{subfigure}
  \caption{Flight patterns of the fleet for the validation measurements around the measurement mast array with wind rose from the reference sensor system of the corresponding flights. The dashed blue lines represent the alignments of the drones relative to each other and to the sensors on the masts.}
  \label{fig:patterns_windroses}
\end{figure}

We conducted the validation flights in measurement campaigns in July 2024 and March 2025. We operated the UASs in two fleet patterns: The first was applied in the 2025 campaign and consists of 2x2 UASs, with two UASs equipped with 5.0$^{\prime\prime}$ propellers and two with 5.5$^{\prime\prime}$ propellers. The two drones with the same setup respectively fly at altitudes of 25~m and 100~m, at a perpendicular distance of 50~m from the measurement masts (see Fig.~\ref{fig:patterns_windroses_a}). We swapped the positions of the upper and lower UASs in the pairs, the positions of the UAS pairs, and positioned the UAS pattern both upstream and downstream of the measurement masts to rule out local or UAS-specific effects. In total, we conducted 13 fleet flights, resulting in 48 usable time series averaging 9.8 minutes. 
The second flight pattern was applied in the 2024 campaign and is a box-shaped pattern in which 10 drones, equipped with 5.5$^{\prime\prime}$ propellers, are positioned at a perpendicular distance of 50~m to the measurement mast array to the west and east of it. UASs are positioned at 25~m and 90~m on the northern and southern masts, and only at 90~m on the middle mast due to the mast guy wires (see Fig.~\ref{fig:patterns_windroses_b}). We conducted 17 fleet flights, resulting in 142 time series averaging 10.4 minutes.

 \section{Method}

 \subsection{Turbulence parameters}
 We examine various turbulence parameters using the time series of the measured wind speed $V_{\mathrm{g}}=V$ and its 3D components $u$, $v$, and $w$:
 In principle, the notation
 \begin{equation}\label{eq:decomposition} 
 s(t)=\overline{s}+s'(t)
 \end{equation} 
 for the Reynolds decomposition of the fluctuating part $s'(t)$ and the mean part $\overline{s}$ defined as
 \begin{equation}\label{eq:mean_s}
 \overline{s}=\frac{1}{T}\int_{0}^{T} s(t)\,\mathrm{dt} 
 \end{equation}
 applies of the time series $s(t)$ with a total time $T$. As a measure of the absolute intensity of the velocity fluctuations we use the variance $\sigma^2$. For $V$ (and analogously for its' components $u$, $v$, and $w$) with a length of $N$, it is defined as
\begin{equation}\label{eq:var-V-sample}
{\sigma_V^2}=\frac{1}{N-1}\sum_{i=1}^{N}\big(V_i-\overline{V}\big)^2 ~.
\end{equation}
We use the covariance $C$ as a measure of the spatial correlation of the flow measured at two spatially distributed positions. For the time series of wind speeds $V_1$ and $V_2$ with the number of sample points $N$ measured at two measurement points, it is calculated as 
\begin{equation}\label{eq:cov-V1V2-sample}
C_{V_1,V_2}=\frac{1}{N-1}\sum_{i=1}^{N}\big(V_{1,i}-\overline{V_1}\big)\big(V_{2,i}-\overline{V_2}\big) ~.
\end{equation}
To quantify the momentum transport, we use the covariance $C_{u,w}$ between the individual velocity components $u$ and $w$ (and analogous $C_{u,v}$ and $C_{w,v}$) for each of all $N$ time steps of the time series:
\begin{align} \label{cov_components}
C_{u,w}=\frac{1}{N-1}\sum_{i=1}^{N}\big(u_{i}-\overline{u}\big)\big(w_{i}-\overline{w}\big)
\end{align}
For the quantification of the energetic distribution of turbulence across different scales, we use the one-sided power spectral density (PSD). The PSD $S$ of the wind speed $V$ as a function of frequency $f$ is calculated using a Fourier transform. The measured $V$ is given in the sampling frequency $f_{\mathrm{s}}$ as time series $V_{\mathrm{i}}$ with \(i = 0,\dots,N-1\). The discrete Fourier transform $\hat{V}_k $ is
\begin{equation}
\hat{V}_k 
= \sum_{i=0}^{N-1}V_i \,\mathrm{e}^{-\,\mathrm{i}\,2\pi \frac{k i}{N}},
\qquad
k = 0,\dots,N-1.
\end{equation}
To obtain a one-sided PSD according to Nyquist's theorem, we define the highest Fourier index as $N_{\mathrm h} = \left\lfloor \frac{N}{2} \right\rfloor$. The corresponding PSD \(S_k\) for each frequency index $k$ then is
\begin{equation}
S_k =
\begin{cases}
\displaystyle
\frac{1}{N f_s}
\left|
 \hat{V}_k 
\right|^2,
& k = 0 \ \text{or}\ k = N_{\mathrm h},
\\[2ex]
\displaystyle
\frac{2}{N f_s}
\left|
\hat{V}_k 
\right|^2,
& 1 \le k \le N_{\mathrm h}-1,
\end{cases}
\qquad
k = 0,\dots,N_{\mathrm h}.
\end{equation}

The corresponding frequency axis of the one-sided spectrum is 
\begin{equation}
f_k = \frac{k}{N} \, f_s,
\qquad
k = 0,\dots,N_{\mathrm h}.
\end{equation}
This gives us $(\boldsymbol{f}, \boldsymbol{S}) 
= \bigl(\{f_k\}_{k=0}^{N_{\mathrm h}}, \{S_k\}_{k=0}^{N_{\mathrm h}}\bigr)$ for $V$.

\subsection{Error metrics}
We use the root mean square error (RMSE) $\varepsilon$ as an absolute measure of the deviation of the UAS measurements from the reference measurements for the number of measurements $N$:
\begin{equation}\label{eq:rmse}
\varepsilon_s = 
\sqrt{
\frac{1}{N}
\sum_{i=1}^{N}
\big( s_{\mathrm{UAS},i} - s_{\mathrm{ref},i} \big)^{2}
}.
\end{equation}
In addition, we use the normalized root mean square mean error (nRMSE) $\varepsilon^{\ast}$ as a relative error measure for quantifying the bias between the UAS measurements and the reference measurements for the number of measurements $N$:

\begin{equation} \label{eq:nrmse}
\varepsilon^{\ast}_s 
= \frac{\displaystyle\sqrt{\frac{1}{N} \sum_{i=1}^{N}\left(s_{\mathrm{UAS},i}-s_{\mathrm{ref},i}\right)^2}}
{\displaystyle \frac{1}{N} \sum_{i=1}^{N} s_{\mathrm{ref},i}}.
\end{equation}

Since covariances are often close to zero, a direct relative error measure such as the nRMSE can only be interpreted to a limited extent for covariances. Instead, we use the Pearson correlation coefficient $r$ to describe the agreement between two measured wind speed components $s_1$ and $s_2$ of the length $N$, and take the RMSE (Eq. \ref{eq:rmse}) between the correlation coefficients derived with the UAS and the reference sensor to get a relative error measure for covariance determination:
\begin{equation} \label{eq:pearsonr}
r_{12} = 
\frac{\displaystyle \sum_{i=1}^N (s_{\mathrm{1},i}- \bar{s}_{\mathrm{1}})(s_{\mathrm{2},i}- \bar{s}_{\mathrm{2}})}
     {\displaystyle 
      \sqrt{\sum_{i=1}^N (s_{\mathrm{1},i} - \bar{s}_{\mathrm{1}})^2}
      \;\sqrt{\sum_{i=1}^N (s_{\mathrm{2},i}- \bar{s}_{\mathrm{2}})^2}
     } \,.
\end{equation}
\subsection{Horizontal calibration}
We perform the calibrations in the wind tunnel for the QAV250 drone equipped with both the 5.0$^{\prime\prime}$ and the 5.5$^{\prime\prime}$ propellers. The reference speed is only available in the geodetic coordinate system. In order to obtain the velocity components in the body-fixed coordinate system so that calibration can be performed using the acceleration data in the body-fixed coordinate system, we use the Euler angles measured by the autopilot to convert the reference velocities into the body-fixed coordinate system. The following assumptions apply to the horizontal calibration flights in the wind tunnel:
\begin{itemize}
 \item The wind direction is solely longitudinal in the main axis of the wind tunnel, cross-wind components are negligible.
 \item The drone maintains its position and its longitudinal axes is perfectly aligned with the direction of the wind.
\end{itemize}
It follows that  $v_{\mathrm{b}} = 0$, $w_{\mathrm{g}} = 0$,  $\psi = \phi = 0$,  and $u_{\mathrm{loc}} = v_{\mathrm{loc}} = w_{\mathrm{loc}} = 0$.

This simplifies $u_{\mathrm{g}}$ for the calibrations to
\begin{equation}\label{eq:u_cal}
 u_{\mathrm{g}} = \cos\theta\,u_{\mathrm{b}}
 + \sin\theta\,w_{\mathrm{b}} 
\end{equation}
and $w_{\mathrm{g}}$ to
\begin{equation}\label{eq:w_cal}
 w_{\mathrm{g}} = 0
 = -\sin\theta\,u_{\mathrm{b}}
 + \cos\theta\,w_{\mathrm{b}}  ~.
\end{equation}
From Eq. \ref{eq:w_cal} it follows that 
\begin{equation} \label{eq:w_body}
    w_{\mathrm{b}} = \tan\theta\,u_{\mathrm{b}}
\end{equation}
which, when substituted into Eq. \ref{eq:u_cal}, yields
\begin{equation}\label{eq:u_cal_body}
 u_{\mathrm{g}} = \frac{u_{\mathrm{b}}}{\cos\theta } \rightarrow   u_{\mathrm{b}} = \cos\theta \,u_{\mathrm{g}}  ~.
\end{equation}
As described by \cite{Kistner2024}, we use the extended Rayleigh drag equations from \cite{Wetz2022} to determine $u_{\mathrm{b}}$ from the logged acceleration data $a_{\mathrm{x}}$ measured in stepwise increased wind speeds and the known mass $m$:
\begin{equation}
u_{\mathrm{b}} =  c_{\mathrm{x}}(ma_{\mathrm{x}})^{b_{\mathrm{x}}}    
\end{equation}
Eq. \ref{eq:u_cal_body} gives us $u_{\mathrm{b}}$, and we obtain the acceleration data $a_{\mathrm{x}}$ from the autopilot. A curve fit can thus be used to determine the calibration coefficients $c_{\mathrm{x}}$ and $b_{\mathrm{x}}$. We also obtain the $w_{\mathrm{b}} $ component in the x-direction (see Eq. \ref{eq:w_body}), which is caused by the horizontal wind and can therefore be determined from the $a_{\mathrm{x}}$ data. 
Analogous to $u_{\mathrm{b}}$, $v_{\mathrm{b}}$ and thus the calibration coefficients $c_{\mathrm{y}}$ and $b_{\mathrm{y}}$ can be determined. Here, the assumption that the drone is perfectly aligned with the wind does not apply; instead, it is assumed to be perfectly aligned at $\pm$90$^{\circ}$ to the wind.  

\begin{figure}[H]
  \centering
  \includegraphics[width=0.45\textwidth]{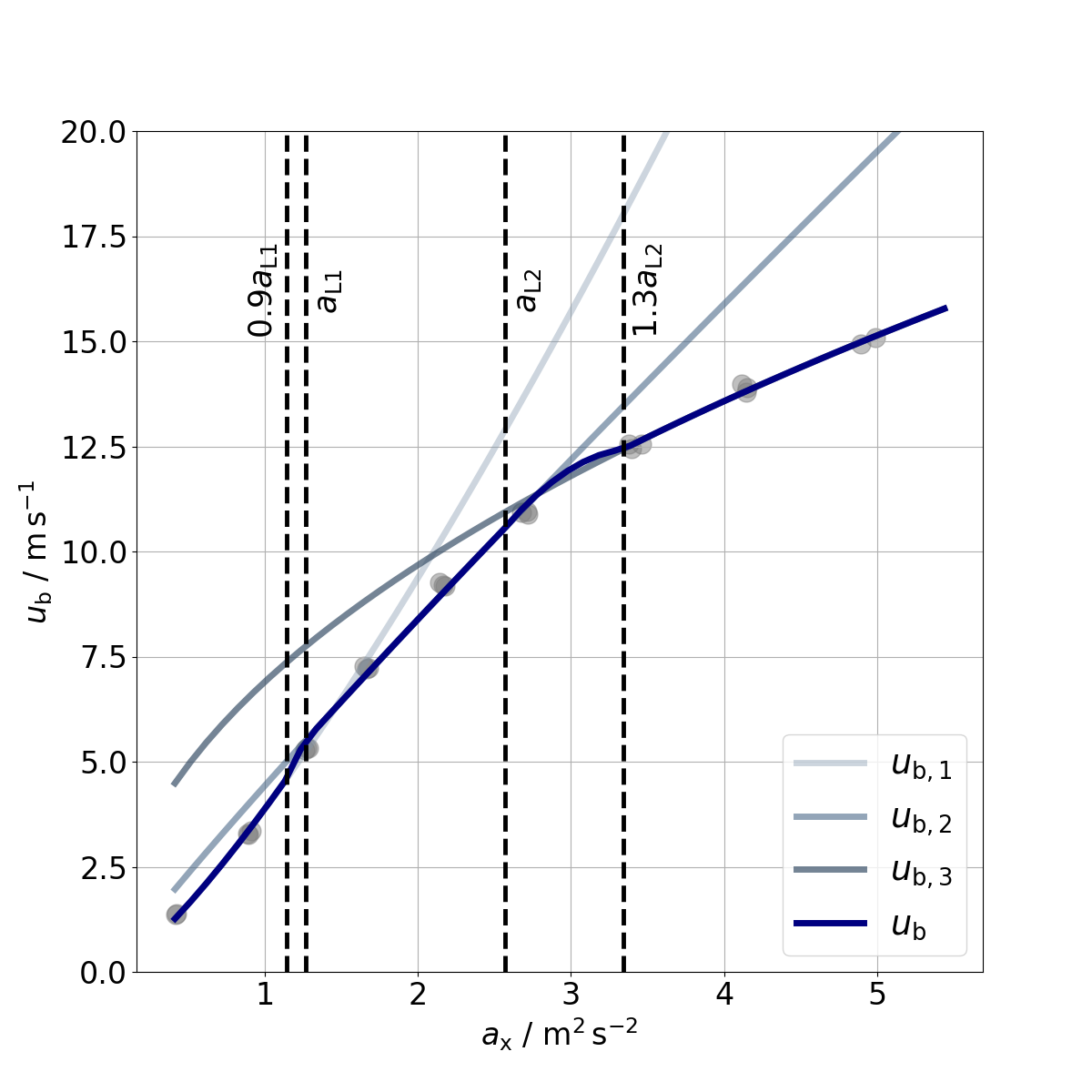} 
  \caption{Calibration curve for the QAV250 drone in the 5.5$^{\prime\prime}$ setup (blue) for determining the longitudinal wind speed in the drone's body-fixed coordinate system $u_{\mathrm{b}}$ based on the acceleration data in the x-direction of the drone's coordinate system $a_{\mathrm{x}}$. The curve (blue) is composed of various calibration curves (grayscale), each of which is fitted for different speed ranges using the data points from calibration flights (gray dots). The dashed black lines indicate the thresholds $a_{\mathrm{L1}}$ and $a_{\mathrm{L2}}$ for the accelerations at which between curves is switched and, in some cases, interpolated between them.}
  \label{fig:interpolated}
\end{figure}

As already described by \cite{Kistner2024}, the adjustment behavior of the drone's pitch angle varies in different speed ranges. We therefore achieve a higher accuracy in wind speed measurements by not determining calibration coefficients $c_{\mathrm{x}}$ and $b_{\mathrm{x}}$, that cover the entire speed range the drone can be operated in, but instead find separate calibration coefficients for low (1), medium (2), and high (3) wind speeds:
\begin{equation}
u_{\mathrm{b},i} =  c_{\mathrm{x},i}(ma_{\mathrm{x}})^{b_{\mathrm{x},i}}    , \qquad i \in \{1,2,3\}.
\end{equation}
For a time series of acceleration data, it is necessary to select the calibration coefficients to be used for converting it into wind speeds for each individual data point. The calibration curve for $u_{\mathrm{b}}$ from $a_{\mathrm{x}}$ is therefore composed of three curves (see Fig.~\ref{fig:interpolated}). These meet at the acceleration values $a_{\mathrm{L1}}$ and $a_{\mathrm{L2}}$, at which the drone's behavior changes in order to counteract the horizontal wind. The first calibration curve transition results from the change from momentum drag to body drag as the dominant drag force with increasing wind speed and thus increasing pitch angle, as also observed by \cite{Pflimlin2010a}. The second calibration curve transition is caused by the pitch angle exceeding the effective blade angle of the propellers, so that the propeller's passive lift force effect caused by the flows around them becomes directed with the wind and therefore a drag force. To prevent discontinuities that would lead to erroneous artifacts in the processing of wind speed, we interpolate between the calibration curves in their transition areas:
\begin{equation}
\begin{aligned}
N := \lvert u_{\mathrm{b,2}}\rvert,\qquad I := \{0,1,\dots,N-1\}, \qquad \forall i\in I: \qquad 
\\
\quad \mathrm{u_{\mathrm{b}}}_i =
\begin{cases} 
\mathrm{u_{\mathrm{b,1}}}_i & a_{\mathrm{x}} \le 0.9\,a_{\mathrm{L1}} \\[6pt]
(1-\frac{a_{\mathrm{x}} - 0.9\,a_{\mathrm{L1}}}{\,0.1\,a_{\mathrm{L1}}\,})\,\mathrm{u_{\mathrm{b,1}}}_i + \frac{a_{\mathrm{x}} - 0.9\,a_{\mathrm{L1}}}{\,0.1\,a_{\mathrm{L1}}\,}\,\mathrm{u_{\mathrm{b,2}}}_i & 0.9\,a_{\mathrm{L1}} < a_{\mathrm{x}} \le a_{\mathrm{L1}} \\[6pt]
\mathrm{u_{\mathrm{b,2}}}_i & a_{\mathrm{L1}} < a_{\mathrm{x}} \le a_{\mathrm{L2}} \\[6pt]
(1-\frac{a_{\mathrm{x}} - a_{\mathrm{L2}}}{\,0.3\,a_{\mathrm{L2}}\,})\,\mathrm{u_{\mathrm{b,2}}}_i + \frac{a_{\mathrm{x}} - a_{\mathrm{L2}}}{\,0.3\,a_{\mathrm{L2}}\,}\,\mathrm{u_{\mathrm{b,3}}}_i& a_{\mathrm{L2}} < a_{\mathrm{x}} \le 1.3\,a_{\mathrm{L2}} \\[6pt]
\mathrm{u_{\mathrm{b,3}}}_i & a_{\mathrm{x}} > 1.3\,a_{\mathrm{L2}}~. 
\end{cases} 
\end{aligned}
\end{equation} \label{eq:cal_ranges}

We only use different calibration ranges to determine the $u_{\mathrm{b}}$ component. For the $v_{\mathrm{b}}$ component, one set of calibration coefficients is sufficient because we assume small wind speeds in $y$ direction, as the weather vane mode always yaws the drone into the main wind direction. 

\subsection{Vertical calibration}
\begin{figure}[H]
  \centering
  \begin{subfigure}{0.48\textwidth}
    \centering
    \includegraphics[width=\linewidth]{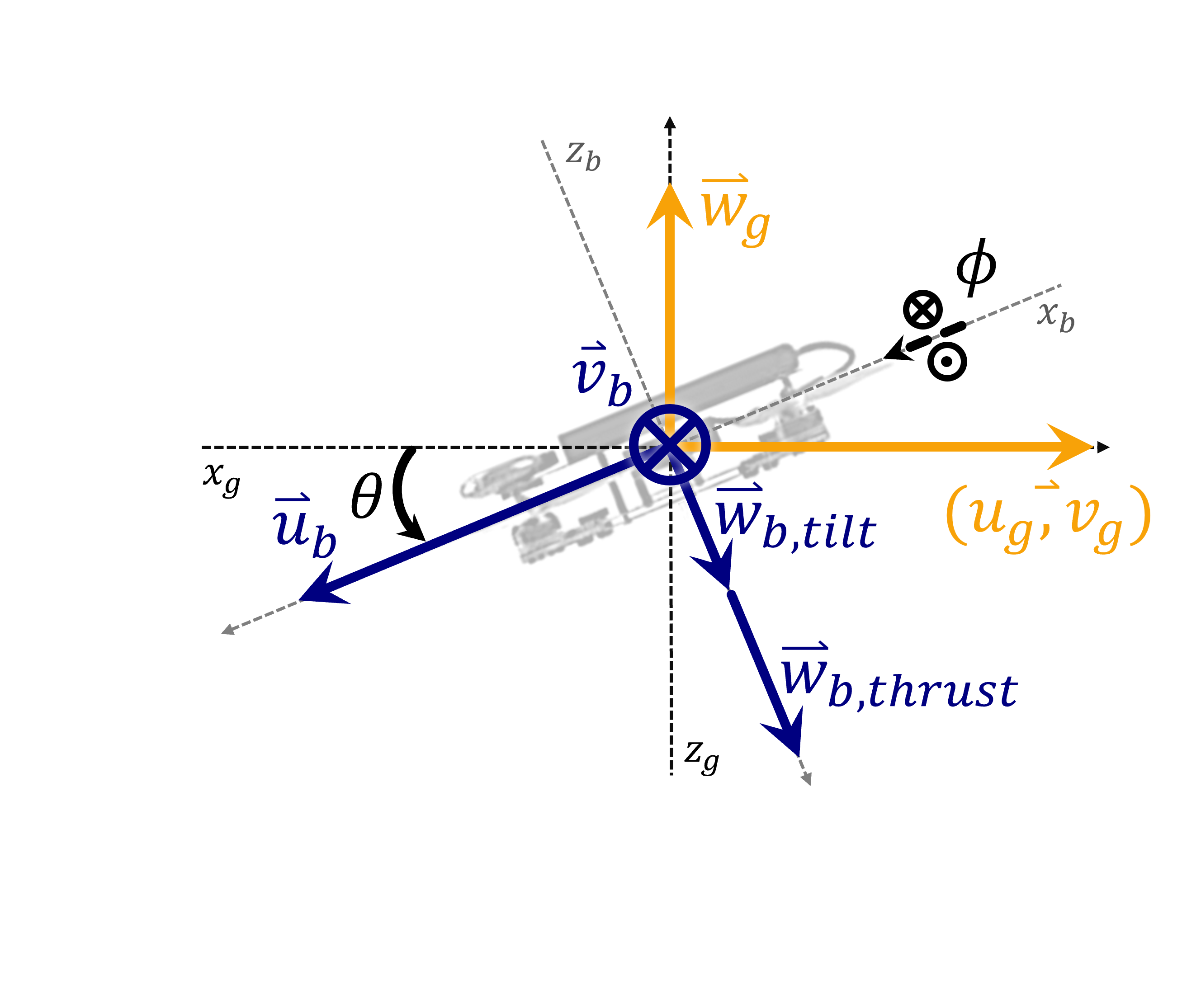}
    \caption{Wind speed components in geodetic (orange) and in the body-fixed coordinate system (blue) of the QAV250 drone and Euler angles (black) in side view (adapted from \citeauthor{Hofmann2025}, \citeyear{Hofmann2025}).}
    \label{fig:speeds}
  \end{subfigure}\hfill
  \begin{subfigure}{0.48\textwidth}
    \centering
    \includegraphics[width=\linewidth]{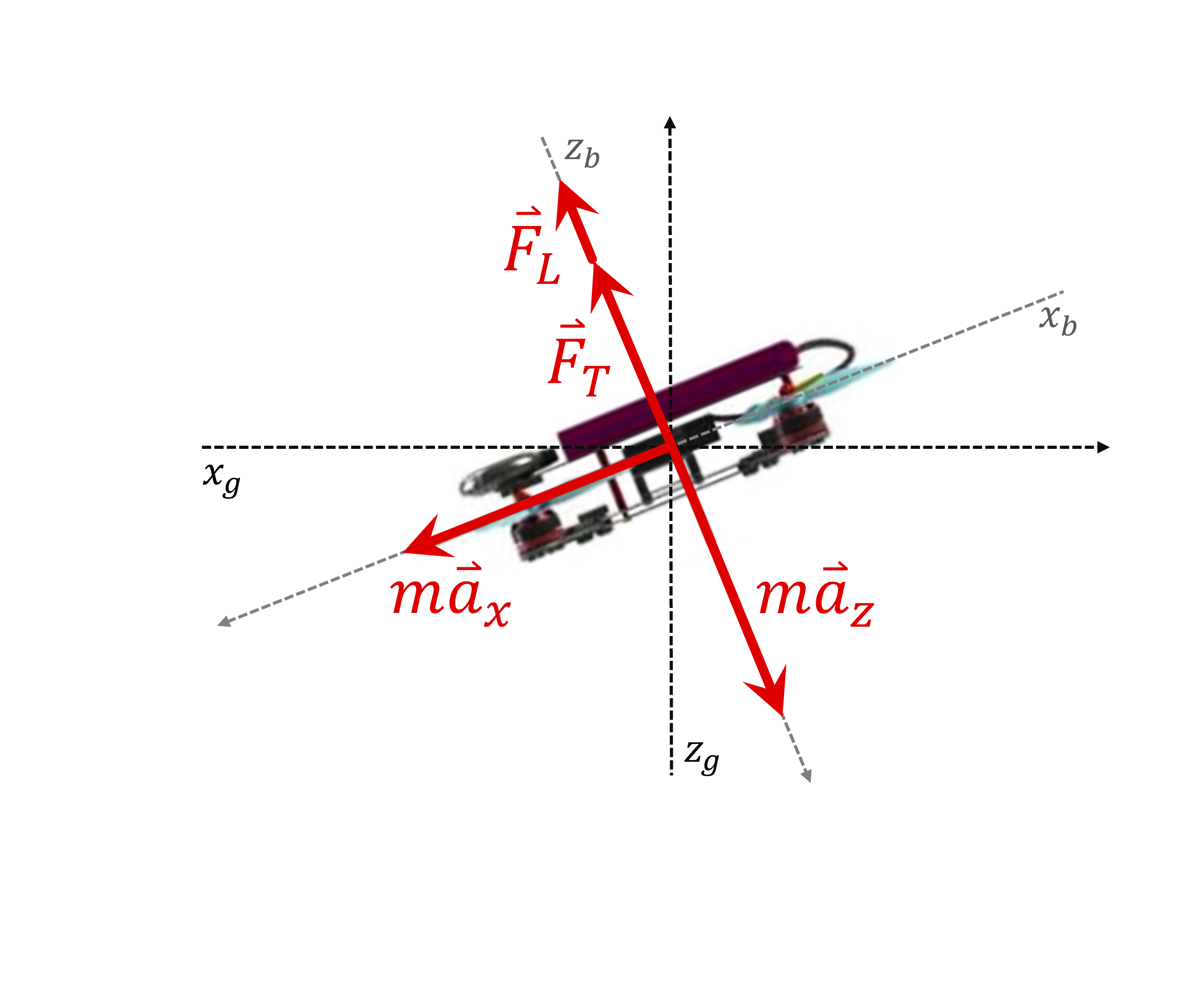}
    \caption{Forces acting on the QAV250 drone in its body-fixed coordinate system in side view (adapted from \citeauthor{Hofmann2025}, \citeyear{Hofmann2025}).\newline}
    \label{fig:forces}
  \end{subfigure}
  \caption{}
\end{figure}

The vertical wind component in the body-fixed coordinate system $w_{\mathrm{b}}$ can be divided into two components  (see Fig.~\ref{fig:speeds}). The first component is caused by the vertical wind, which is counteracted with adjustments of the thrust and we refer to as $w_{\mathrm{b,thrust}}$.  The second component is caused by the horizontal wind and is incorporated into the drone's $z_{\mathrm{b}}$ direction through the tilting of the drone against the wind. We refer to it as $w_{\mathrm{b,tilt}}$:  

\begin{equation} \label{eq:w_components}
w_{\mathrm{b}} = w_{\mathrm{b,thrust}}+w_{\mathrm{b,tilt}} 
\end{equation}
In the conditions for the derivations in the previous subsection we assume $w_{\mathrm{g}} = 0$, which results in $w_{\mathrm{b,thrust}} = 0$. The relationship from Eq. \ref{eq:w_body} therefore describes $w_{\mathrm{b}} =  w_{\mathrm{b,tilt}}$ for the conditions described above. Without these assumptions, the following applies:
\begin{equation} \label{eq:w_body_abs}
    w_{\mathrm{b,tilt}} = \frac{ \tan\theta}{\cos\phi } u_{\mathrm{b}}-\tan\phi\,v_{\mathrm{b}}
\end{equation}

To determine the vertical wind component $w_{\mathrm{g}}$, we use the method introduced by \cite{Wildmann2022} based on the thrust of the drone. For this purpose, the force balance in the $z$ direction of the drone's body-fixed coordinate system is used:
\begin{equation} \label{sum_F_z}
    F_{\mathrm{z}} = \sum F_{\mathrm{z,b}} = ma_{\mathrm{z}} - (F_{\mathrm{L}} + F_{\mathrm{T}})
\end{equation}
with the thrust force $F_{\mathrm{T}}$ generated by the drone's rotors and the aerodynamic lift force $F_{\mathrm{L}}$ resulting from the horizontal flow around the drone (see Fig.~\ref{fig:forces}). 

\begin{figure}[H]
  \centering
  \begin{subfigure}{0.48\textwidth}
    \centering
    \includegraphics[width=\linewidth]{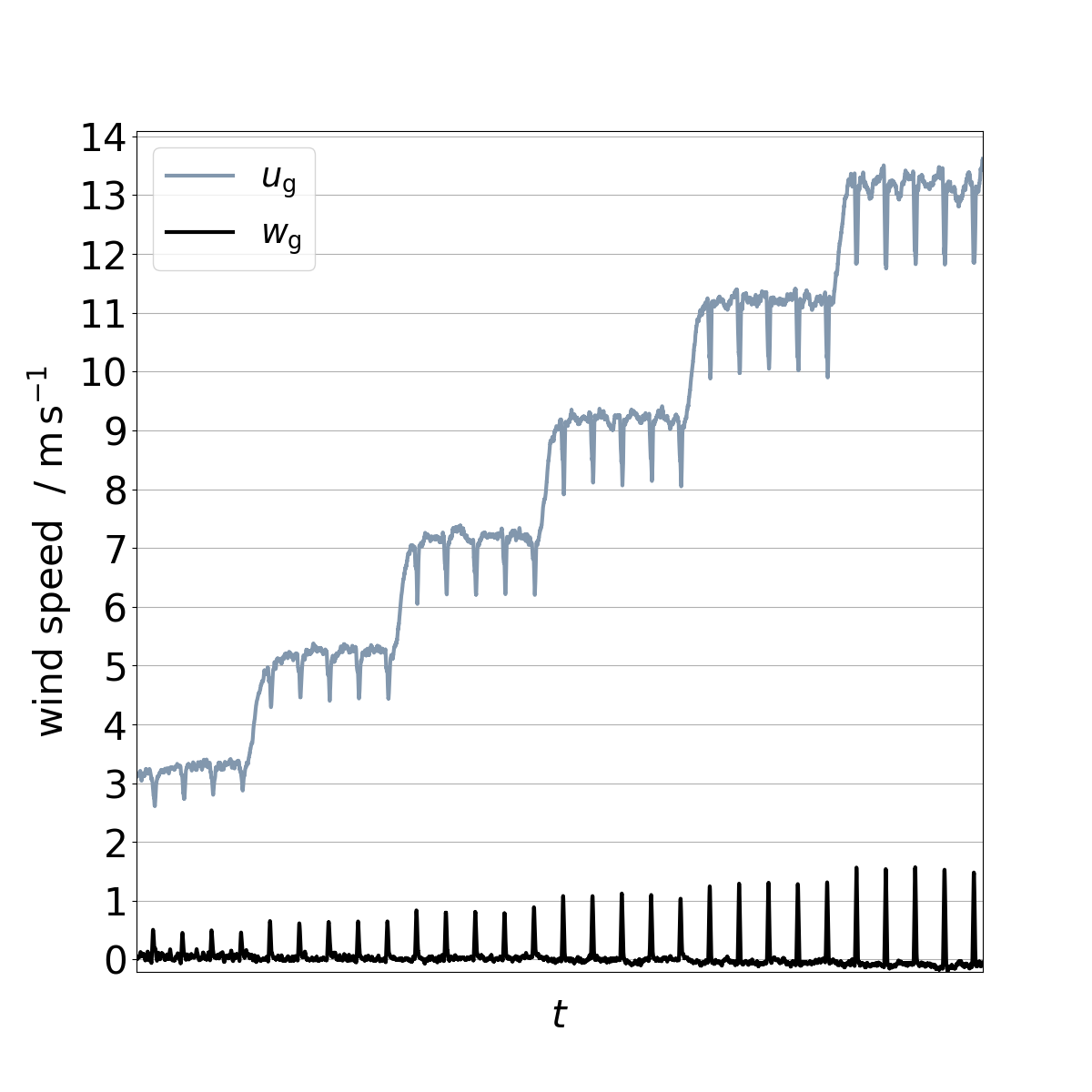}
    \caption{Time series of the velocity components $u_{\mathrm{g}}$ (black) and $w_{\mathrm{g}}$ (gray) measured with the multi hole probe. \newline}
    \label{fig:ts_cal_geo}
  \end{subfigure}\hfill
  \begin{subfigure}{0.48\textwidth}
    \centering
    \includegraphics[width=\linewidth]{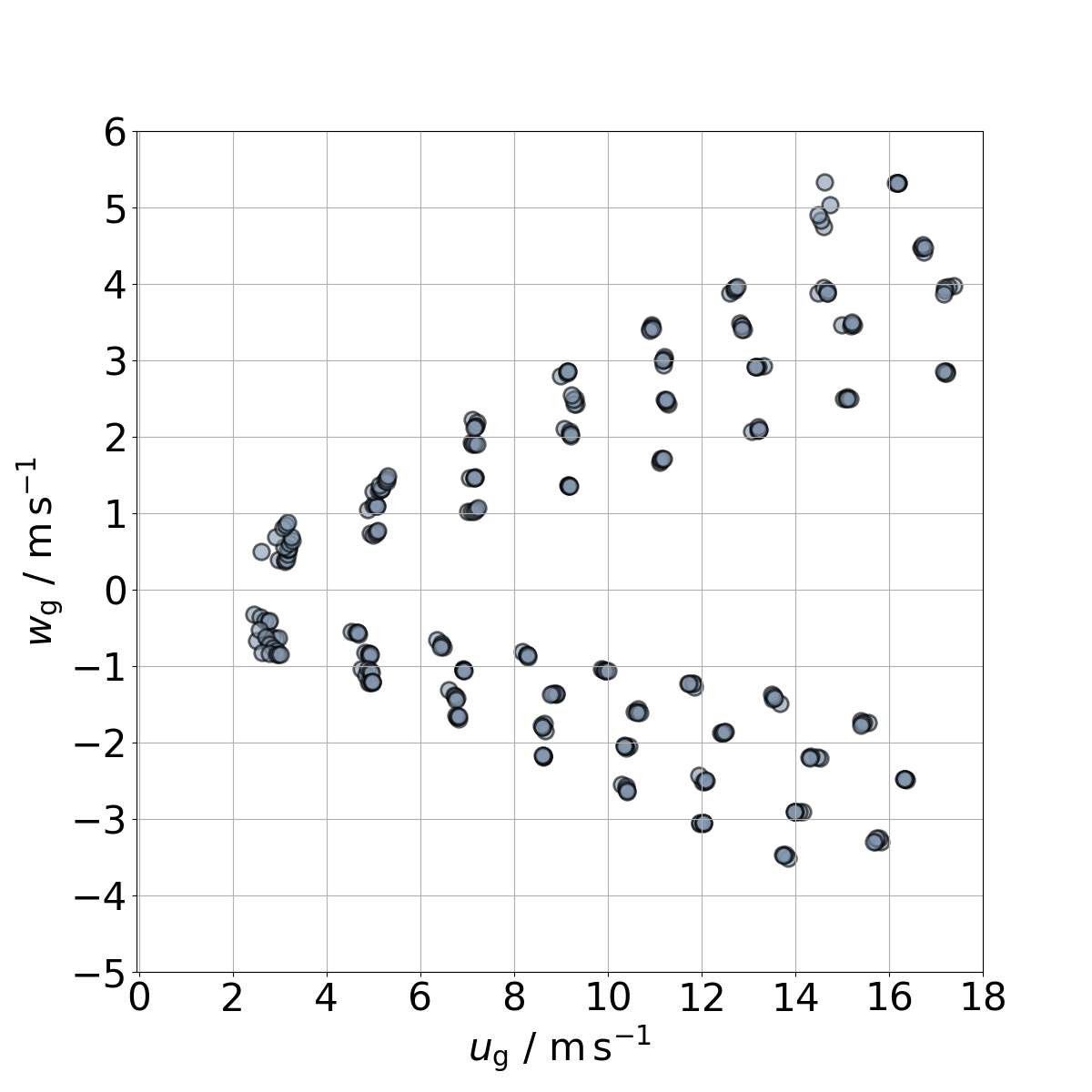}
    \caption{Vertical wind component $w_{\mathrm{g}}$ at the horizontal wind speed component $u_{\mathrm{g}}$ in the geodetic coordinate system used for calibration.}
    \label{fig:scatter_cal_geo}
  \end{subfigure}
  \caption{Reference speeds for calibration of the vertical wind component measurement.}
  \label{fig:reference_speeds_geo}
\end{figure}

To calculate the thrust force, we apply empirical approaches analogous to \cite{Wildmann2022}: We calculate the thrust force using the rotational speed $\omega$ of the propeller. We cannot obtain this directly from the avionics data, but instead calculate it using the pulse-width modulation signals sent to the motors, which we correct by the effective motor voltage. Both quantities are logged as avionics data by the autopilot.
To determine $F_{\mathrm{L}}$, we again make use of the set flow conditions in the wind tunnel: 
\begin{equation} \label{eq:force_balance}
    w_{\mathrm{g}} = 0 \rightarrow F_{\mathrm{z}} = \sum F_{\mathrm{z,i}} = 0 = ma_{\mathrm{z}} - (F_{\mathrm{L}} + F_{\mathrm{T}}) \\
    \rightarrow F_{\mathrm{L}} = ma_{\mathrm{z}} - F_{\mathrm{T}}
\end{equation}
Thereby, $F_{\mathrm{L}}$ consists of the two components resulting from the flows longitudinal and lateral to the drone:
 \begin{equation}
     F_{\mathrm{L}} = \mathrm{sgn(}  F_{\mathrm{L,x}} + F_{\mathrm{L,y}}\mathrm{)} \sqrt{F_{\mathrm{L,x}}^2 + F_{\mathrm{L,y}}^2} 
 \end{equation}
We obtain the acceleration $a_{\mathrm{z}}$ from the autopilot, and we determine $F_{\mathrm{T}}$ based on $\omega$. This allows us to determine $F_{\mathrm{L}}$ empirically as a function of the force $F_{\mathrm{x,b}} =  ma_{\mathrm{x}} $ caused by the horizontal wind acting along the x-axis in the drone's body-fixed coordinate system (cf. \citeauthor{Wildmann2022}, \citeyear{Wildmann2022}). For a best curve fit, in contrast to \cite{Wildmann2022}, who fit a 5th degree polynomial to field data, we set a 2nd degree polynomial for $F_{\mathrm{L,x}}$ and a 3rd degree polynomial for $F_{\mathrm{L,x}}$:
\begin{equation}
     F_{\mathrm{L,x}} = c_{\mathrm{0,x}} + c_{\mathrm{1,x}}a_{\mathrm{x}} + c_{\mathrm{2,x}}a_{\mathrm{x}}^2  ~,
     \end{equation} \begin{equation}
     F_{\mathrm{L,y}} = c_{\mathrm{0,y}} + c_{\mathrm{1,y}}a_{\mathrm{y}} + c_{\mathrm{2,y}}a_{\mathrm{y}}^2 + c_{\mathrm{3,y}}a_{\mathrm{y}}^3
 \end{equation}
where $ c_{\mathrm{0,x}}$ and  $c_{\mathrm{0,y}}$ are offsets that need to be determined individually, analogous to the horizontal offsets \citep{Wetz2022}. We know $m$ and obtain $a_{\mathrm{x}}$, $a_{\mathrm{y}}$, and $a_{\mathrm{z}}$ directly, and $\omega$ indirectly from the avionics data, so we know $F_{\mathrm{T}}=f(\omega)$, $F_{\mathrm{L}}=f(a_{\mathrm{x}}, a_{\mathrm{y}})$ and $F_{\mathrm{z,b}} = ma_{\mathrm{z}}$, which is why we know the force balance from Eq. \ref{sum_F_z}. Based on this, we perform a calibration to determine $w_{\mathrm{g}}$ in the wind tunnel.  
\newline
To generate wind speeds in the $z_{\mathrm{g}}$ direction, we use the capabilities of the active grid by using the flaps to deflect the horizontal wind into the vertical. This is not a constant adjustment, but only possible in the form of vertical gusts. We use different deflection amplitudes of the active grid's flaps to generate vertical gusts of varying strength. Specifically, these are deflections of $-25^{\circ}, -20^{\circ}, ..., 20^{\circ}, 25^{\circ}$. To ensure that we can reliably measure different $w_{\mathrm{g}}$ values for different $u_{\mathrm{g}}$ values (and vice versa), we gradually increase the horizontal wind speed for equal amplitude deflections (see Fig.~\ref{fig:ts_cal_geo}). The resulting combinations of values for $u_{\mathrm{g}}$ and $w_{\mathrm{g}}$ can be seen in Fig.~\ref{fig:scatter_cal_geo}.

For the calibration flights in the wind tunnel to determine the vertical wind component, the assumptions from chapter 3.b. no longer apply; only $\psi = 0$ is still valid due to the weather vane mode. Thus, the following applies to the measured reference velocity in the vertical direction $w_{\mathrm{g}}$:
\begin{equation}
     w_{\mathrm{g}} = \sin\theta\,u_{\mathrm{b}}  - \cos\theta\sin\phi\,v_{\mathrm{b}}  - \cos\theta\cos\phi\,w_{\mathrm{b}}+w_{\mathrm{loc}}~.
\end{equation}
After solving for $w_{\mathrm{b}}$, the following applies:
\begin{equation} \label{eq:w_body_cal}
     w_{\mathrm{b}} = \frac{-w_{\mathrm{g}} +\sin\theta\,u_{\mathrm{b}}  - \cos\theta\sin\phi\,v_{\mathrm{b}}+w_{\mathrm{loc}}} { \cos\theta\cos\phi} ~.
\end{equation}
When equations \ref{eq:w_body_abs} and \ref{eq:w_body_cal} are substituted into Eq. \ref{eq:w_components} and solved for $w_{\mathrm{b,thrust}}$, the result is
\
\begin{align} \label{eq:w_body_cal_full}
     w_{\mathrm{b,thrust}}
     = \frac{-w_{\mathrm{g}} +\sin\theta\,u_{\mathrm{b}}  - \cos\theta\sin\phi\,v_{\mathrm{b}}+w_{\mathrm{loc}}} { \cos\theta\cos\phi} -\frac{ \tan\theta}{\cos\phi } u_{\mathrm{b}}+\tan\phi\,v_{\mathrm{b}} 
     = \frac{-w_{\mathrm{g}} +w_{\mathrm{loc}}} { \cos\theta\cos\phi} ~.
\end{align}
We obtain $w_{\mathrm{g}}$ from the reference sensors, $\theta$ and $\phi$ directly from the autopilot, and $u_{\mathrm{b}}$ and $v_{\mathrm{b}}$ indirectly via the calculations from chapter 2. This allows curve fitting $w_{\mathrm{b,thrust}}$ over the force balance $ F_{\mathrm{z}}$ from Eq. \ref{eq:force_balance} to find $b_{\mathrm{z}}$ and $c_{\mathrm{z}}$: 
\begin{equation}
w_{\mathrm{b,thrust}} =  c_{\mathrm{z}}F_{\mathrm{z}}^{b_{\mathrm{z}}}~. 
\end{equation}
\cite{Wildmann2022} already described that the upward and downward movement of the quadcopter relative to the air is subject to different drag forces, which is why we also make a case distinction for the vertical direction:
\begin{equation}
w_{\mathrm{b,thrust}} =
\begin{cases}
c_{\mathrm{z}}^{\uparrow} F_{\mathrm{z}}^{b\uparrow}, & F_z \ge 0 \\[6pt]
c_{\mathrm{z}}^{\downarrow} F_{\mathrm{z}}^{b\downarrow}, & F_z < 0
\end{cases}
\end{equation}

This means that the vertical wind speed $w_{\mathrm{b}}$ and the horizontal wind speeds $u_{\mathrm{b}}$ and $v_{\mathrm{b}}$ can be calculated in the body-fixed coordinate system using the acceleration data and Euler angle data from the autopilot. The components of the drone's velocity $u_{\mathrm{loc}},v_{\mathrm{loc}}$ and $w_{\mathrm{loc}}$ are also obtained from the autopilot. This provides all the data needed to solve Eq. \ref{eq:1.3a}, \ref{eq:1.3b} and \ref{eq:1.3c}.

\subsection{Field validation}
During field validation flights, we use a flight pattern in which UASs are equipped with both 5.0$^{\prime\prime}$ and 5.5$^{\prime\prime}$ propellers (see Fig.~\ref{fig:patterns_windroses_a}) to verify the performance of wind tunnel calibrations in three ways: 
\newline
Firstly, we apply both the calibration coefficients and calculation terms from the field calibrations by \cite{Wildmann2022} and those from the wind tunnel calibrations in this study to the avionics data from the drones with the 5.0$^{\prime\prime}$ setup. This allows us to compare the performance of the two calibrations. It should be noted that we can only use flights for this analysis where $V \leq 8~\mathrm{m}~\mathrm{s}^{-1}$ always applies, as this is the restriction for applying the coefficients and calculations from the field calibration. To investigate the measurement capability of turbulence in the vertical wind component, we analyze its variance.
\newline
Secondly, we compare the variance of the measured $w$ component between the two setups with 5.0$^{\prime\prime}$ and 5.5$^{\prime\prime}$ propellers to check whether the propeller size affects the ability to measure turbulence in the vertical direction. The wind tunnel calibrations are used for this purpose, which means that all flights, including those with $V > 8~\mathrm{m}~\mathrm{s}^{-1}$, can be used.
\newline
Thirdly, we conduct more detailed analyses for the setup with the 5.5$^{\prime\prime}$ propellers, as this is the configuration chosen for future fleet measurements due to the longer flight and measurement duration. During flights with the box-shaped flight pattern (see Fig.~\ref{fig:patterns_windroses_b}), 10 drones are in the air at the same time. This provides more data points and allows for multi-point comparisons at different directions in the same airflow.
\newline
 In the box-shaped flight pattern, we analyze the covariances of the measured wind speeds of UASs located laterally or vertically adjacent with respect to the main wind direction in comparison to their assigned reference sensors. 
Furthermore, we compare the measured PSD of the vertical wind component derived from the drones with the reference sensors, as this is important for turbulence analyses based on the correlation of signals with $w$, such as the above-mentioned covariances between measurement points. 
Since the vertical wind component in the drone's body-fixed coordinates contributes to all three geodetic wind speed components according to the transformation equations (Eq. \ref{eq:1.3a}-\ref{eq:1.3c}), we check the turbulence measurement capability for all three wind speed components using the variance of the individual components and the covariance between the components.
 
 \section{Results}
\begin{figure}[H]
  \centering
  \begin{subfigure}{0.48\textwidth}
    \centering
    \includegraphics[width=\linewidth]{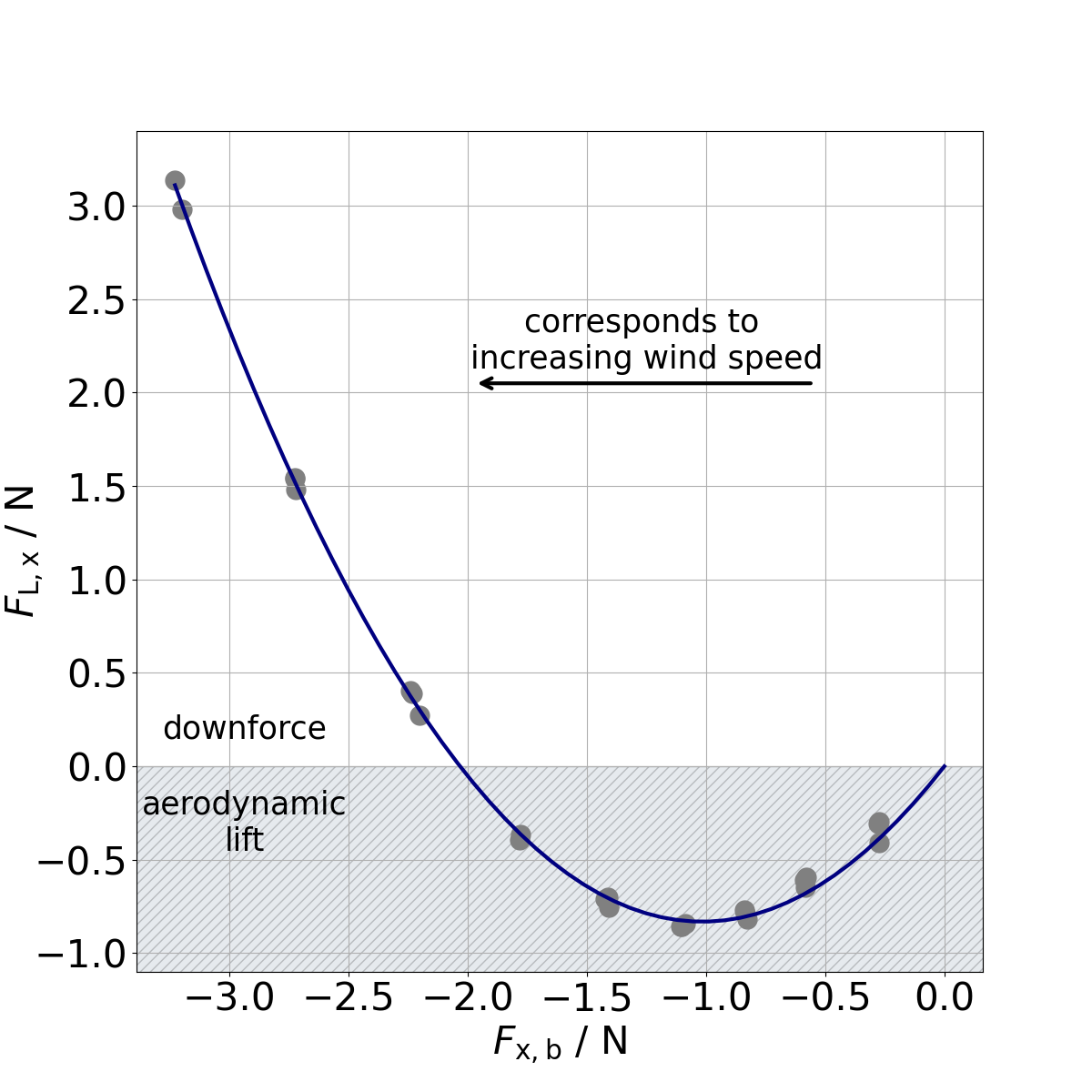}
    \caption{Averaged lift force $F_{\mathrm{L,x}}$ over the force $F_{\mathrm{x,b}}$ caused by the horizontal wind acting on the drone in the 5.5$^{\prime\prime}$ setup. The blue curve represents the curve fit through the measurement data points. \newline }
    \label{fig:lift_scatter}
  \end{subfigure}\hfill
  \begin{subfigure}{0.48\textwidth}
    \centering
    \includegraphics[width=\linewidth]{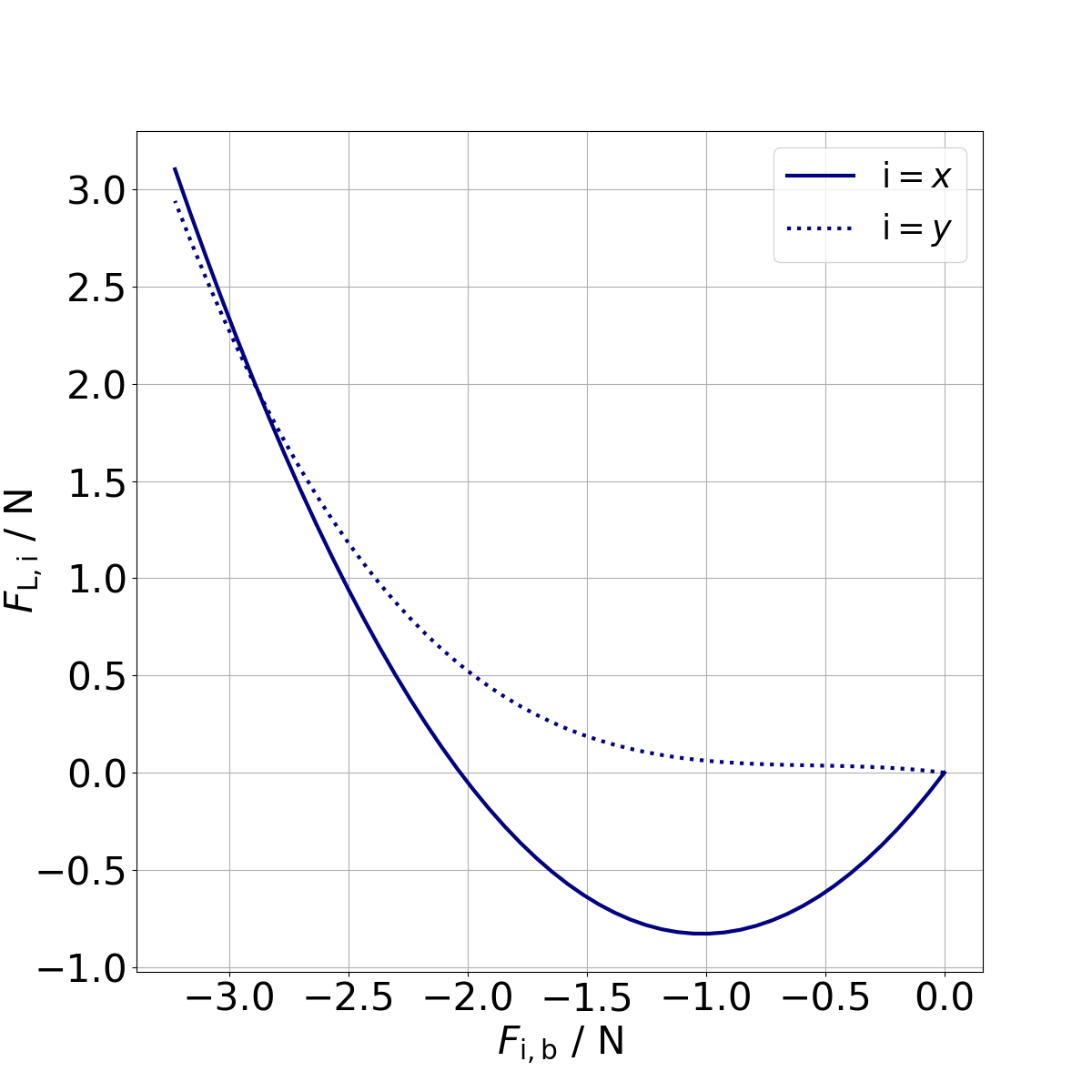}
    \caption{Averaged lift forces $F_{\mathrm{L},i}$ over the force $F_{i,\mathrm{b}}$ caused by the horizontal winds. The solid curve represents the lift force caused by horizontal wind in the longitudinal direction $F_{\mathrm{L,x}}$, the dashed line represents the lift force $F_{\mathrm{L,y}}$ caused by horizontal wind in the lateral direction of the drone.}
    \label{fig:lift_curve_comparison_x_y}
  \end{subfigure}
  \label{fig:reference_speeds_geo}
  \begin{subfigure}{0.48\textwidth}
   \centering
    \includegraphics[width=\linewidth]{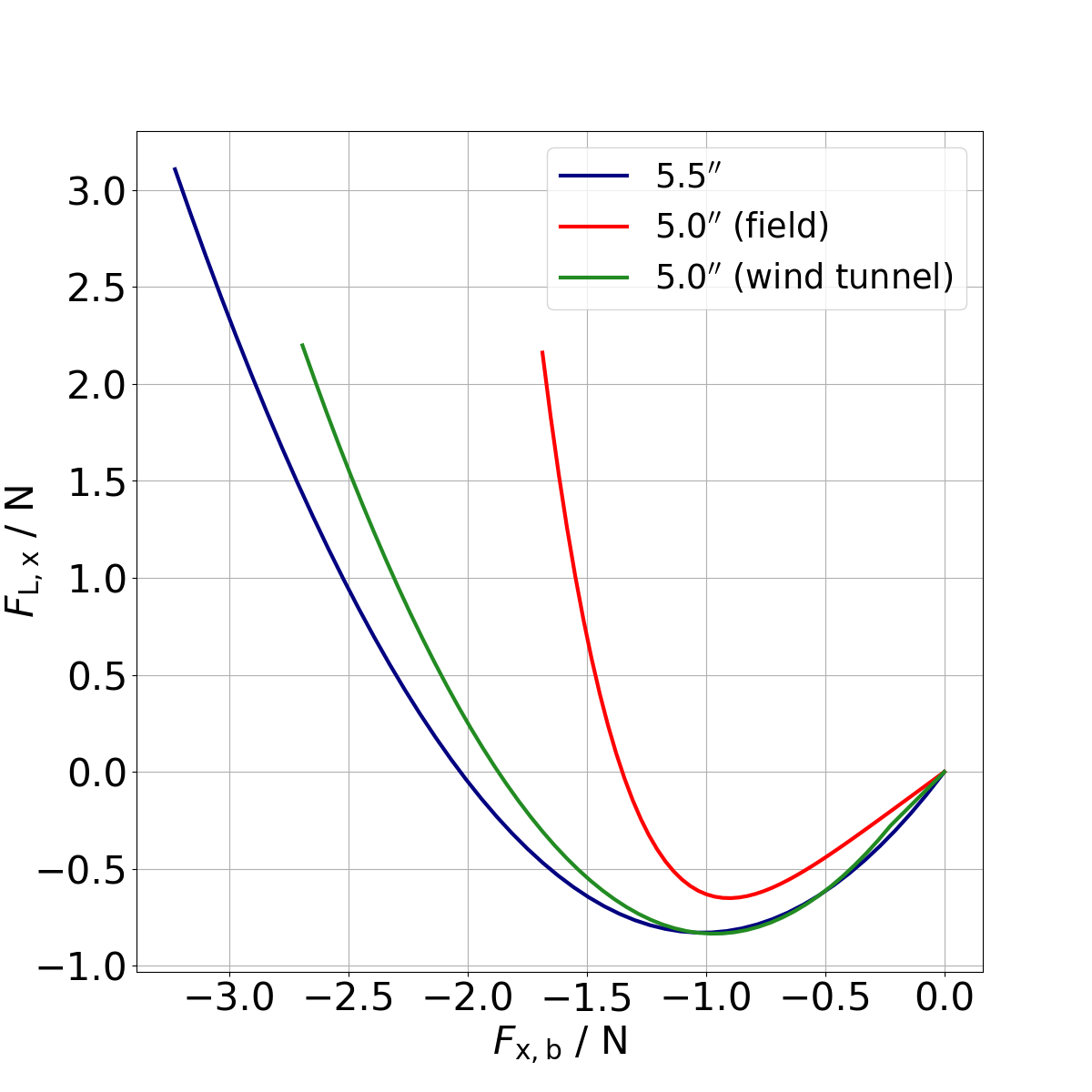}
    \caption{Comparison of lift curves from field calibration (red) by \cite{Wildmann2022}, wind tunnel calibration with the same hardware setup (green), and wind tunnel calibration with the new setup (blue).}
    \label{fig:lift_curve_comparison}
    
  \end{subfigure}
  
  \label{fig:lift_curves}
  \caption{}
\end{figure}

Wind tunnel calibration for measuring horizontal wind enabled the corresponding calibration coefficients to be determined, which are summarized in Table \ref{tab:hor_coeff}.
\begin{table}[H] 
\caption{Calibration coefficients for horizontal wind measurement}\label{tab:hor_coeff}
\centering
\begin{tabular}{c|c|crrrrrrrrrrrrrcrc}
\hline\hline
coefficient &  5.0$^{\prime\prime}$ &  5.5$^{\prime\prime}$   \\
\hline
$a_\mathrm{L1}$ & 1.35 & 1.27  \\
$a_\mathrm{L2}$ & 2.50 & 2.57  \\  
$b_{\mathrm{x,1}}$  & 1.145 &  1.275  \\ 
$c_{\mathrm{x,1}}$ & 7.052 & 6.636  \\ 

$b_{\mathrm{x,2}}$  & 0.842 &  0.923  \\ 
$c_{\mathrm{x,2}}$  & 6.980 & 6.531 \\ 

$b_{\mathrm{x,3}}$  & 0.517 & 0.489   \\ 
$c_{\mathrm{x,3}}$   & 8.513 &  8.477  \\ 
$b_{\mathrm{y}}$ & 0.701  & 0.792  \\ 
$c_{\mathrm{y}}$  & 6.104 & 5.691  \\ 
\hline\hline
\end{tabular}
\end{table}
The wind tunnel measurements and the corresponding boundary conditions enable us to determine the coefficients for defined lift curves up to horizontal wind speeds of approximately 18$~\mathrm{m~s^{-1}}$, which are shown in Fig.~\ref{fig:lift_scatter}. This means that there is no maximum wind speed beyond the flight range limits of the UAS at which vertical wind measurement could no longer be performed.
Note that in the body-fixed coordinate system (see Fig.~\ref{fig:coordinate_system}) a more negative $F_{\mathrm{x,b}}$ corresponds to a higher horizontal wind speed, and a more negative $F_{\mathrm{L,x}}$ is a stronger force lifting the drone upwards. The polynomial for the curve fit of the lift in the $x$ direction is of the second order, while the polynomial for the lift in the $y$ direction is of the third order (cf. Fig.~\ref{fig:lift_curve_comparison_x_y}). 

The comparison of the lift curves between the 5.5$^{\prime\prime}$ and the 5.0$^{\prime\prime}$ propeller setups determined in the wind tunnel experiments, as well as the lift curve by \cite{Wildmann2022} of the 5.0$^{\prime\prime}$ propeller setup determined in field measurements, is shown in Fig.~\ref{fig:lift_curve_comparison}.

\begin{figure}[H]
  \centering
  \begin{subfigure}{0.48\textwidth}
    \centering
    \includegraphics[width=\linewidth]{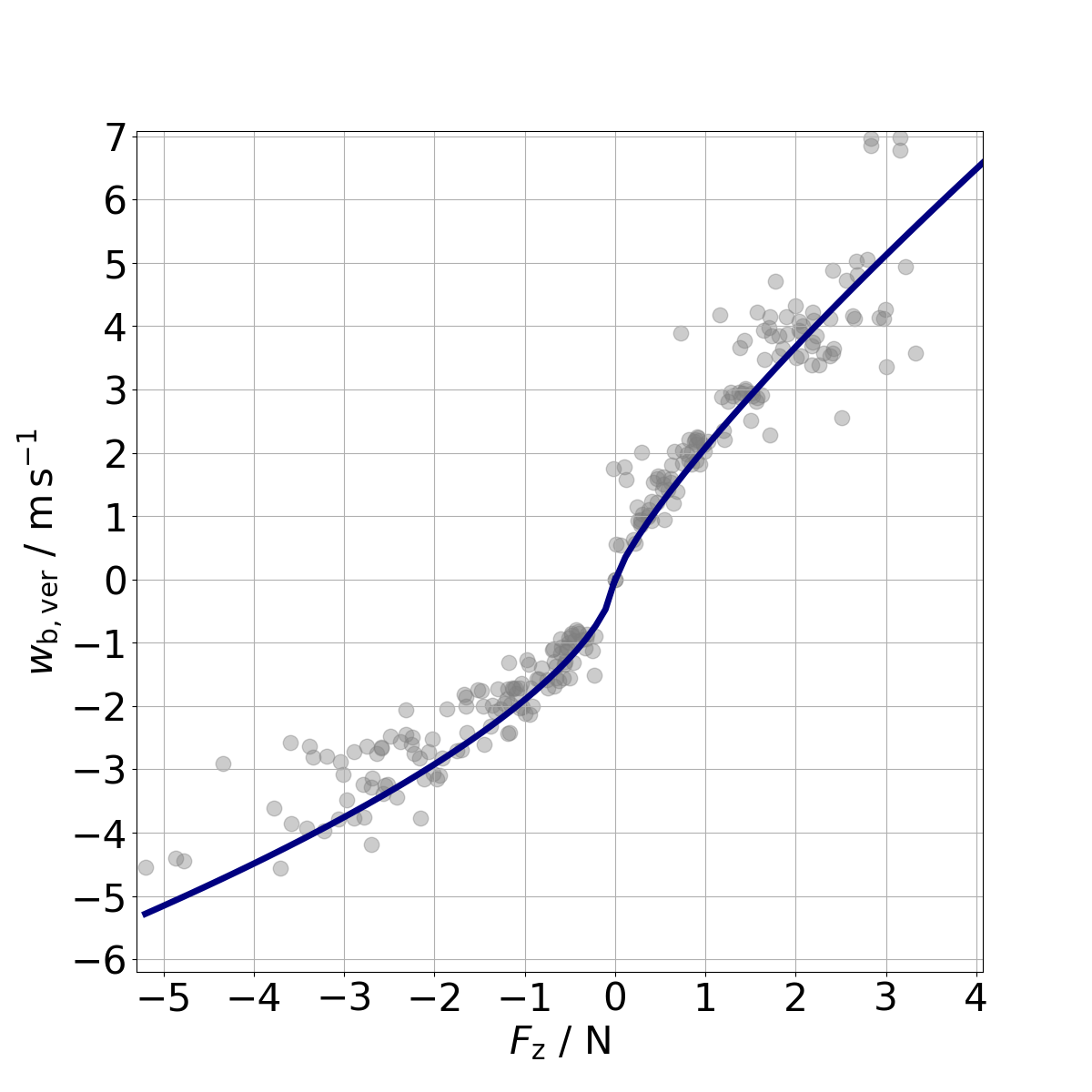}
    \caption{Fitted speed curve for the conversion of the force balance $F_{\mathrm{z}}$ vertically acting on the drone in the 5.5$^{\prime\prime}$ setup to the vertical wind speed $w_{\mathrm{b,thrust}}$ in the body-fixed coordinate system. \newline}
  \label{fig:cal_curve_w}
  \end{subfigure}\hfill
  \begin{subfigure}{0.48\textwidth}
    \centering
    \includegraphics[width=\linewidth]{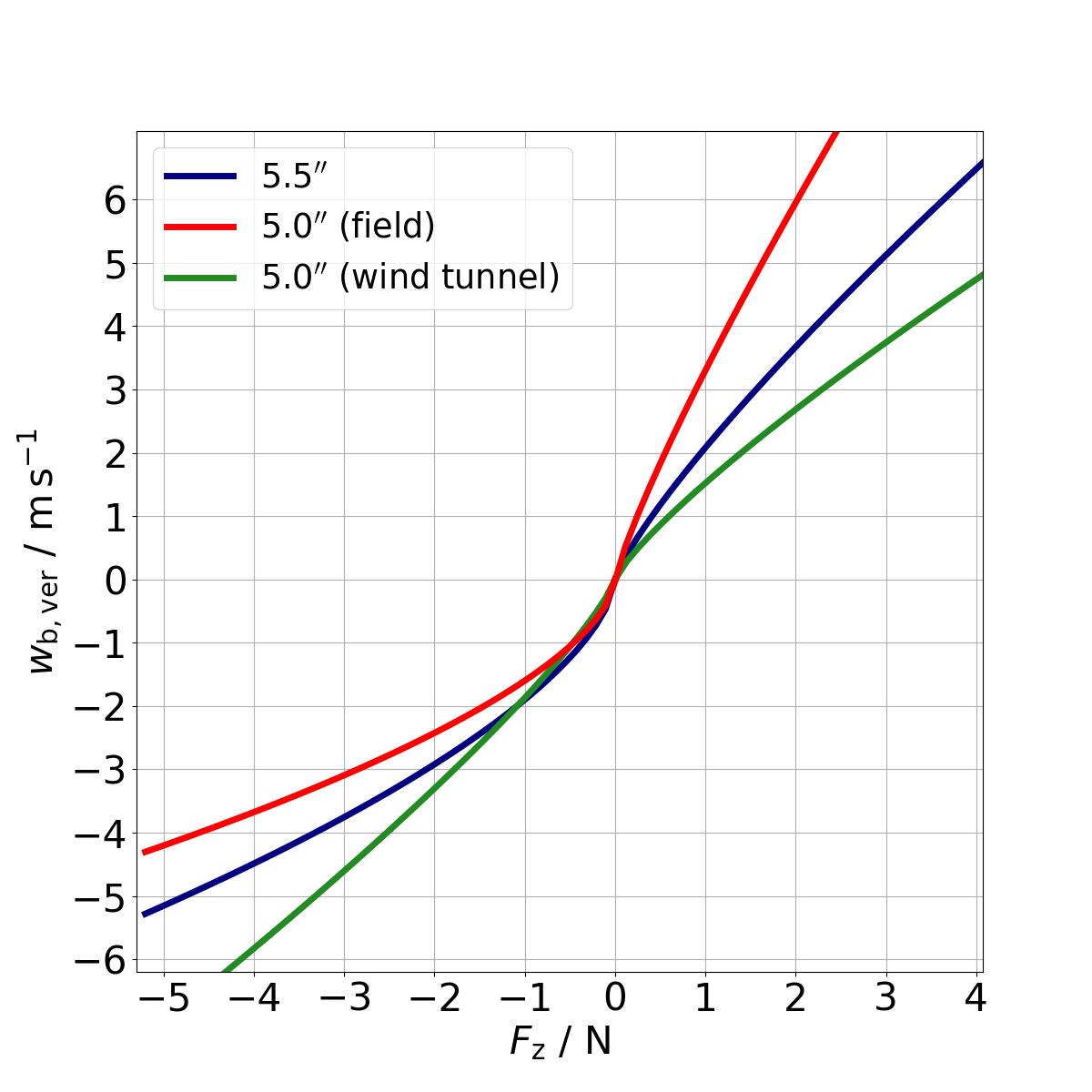}
    \caption{Comparison of speed curves from field calibration (red) by \cite{Wildmann2022}, wind tunnel calibration with the same hardware setup (green), and wind tunnel calibration with the new setup (blue).}
  \label{fig:cal_curve_w_comparison}
  \end{subfigure}
  \caption{}
\end{figure}

The calibration of the measurement of the $w$ component via $F_{\mathrm{z}}$ (see Fig.~\ref{fig:cal_curve_w}) based on this determines the coefficients for measuring the vertical wind, which are listed in Table \ref{tab:ver_coeff}.
\begin{table}[t]
\caption{Calibration coefficients for vertical wind measurement}\label{t1} \label{tab:ver_coeff}
\begin{center}
\begin{tabular}{c|c|crrrrrrrrrrrrrcrc} 
\hline\hline
coefficient &  5.0$^{\prime\prime}$ &  5.5$^{\prime\prime}$   \\
\hline
 
$b_{\mathrm{z}}^{\uparrow}$  & 0.820 & 0.621   \\ 
$c_{\mathrm{z}}^{\uparrow}$ & 1.87 & 1.90  \\ 

$b_{\mathrm{z}}^{\downarrow}$  & 0.821 & 0.820  \\ 
$c_{\mathrm{z}}^{\downarrow}$ & 1.52 & 2.08  \\ 

$c_{\mathrm{1,x}}$  & 1.9639 & 1.6323   \\ 
$c_{\mathrm{2,x}}$  & 1.0167 & 0.8039   \\ 

$c_{\mathrm{1,y}}$  & 0.3005 & 0.1564   \\ 
$c_{\mathrm{2,y}}$  & 0.1029 & 0.2421   \\ 
$c_{\mathrm{3,y}}$  & -0.0918 & 0.1475   \\ 

\hline\hline
\end{tabular}
\end{center}
\end{table}\textbf{}

\begin{figure}[H]
  \centering
  \begin{subfigure}{0.48\textwidth}
    \centering
    \includegraphics[width=\linewidth]{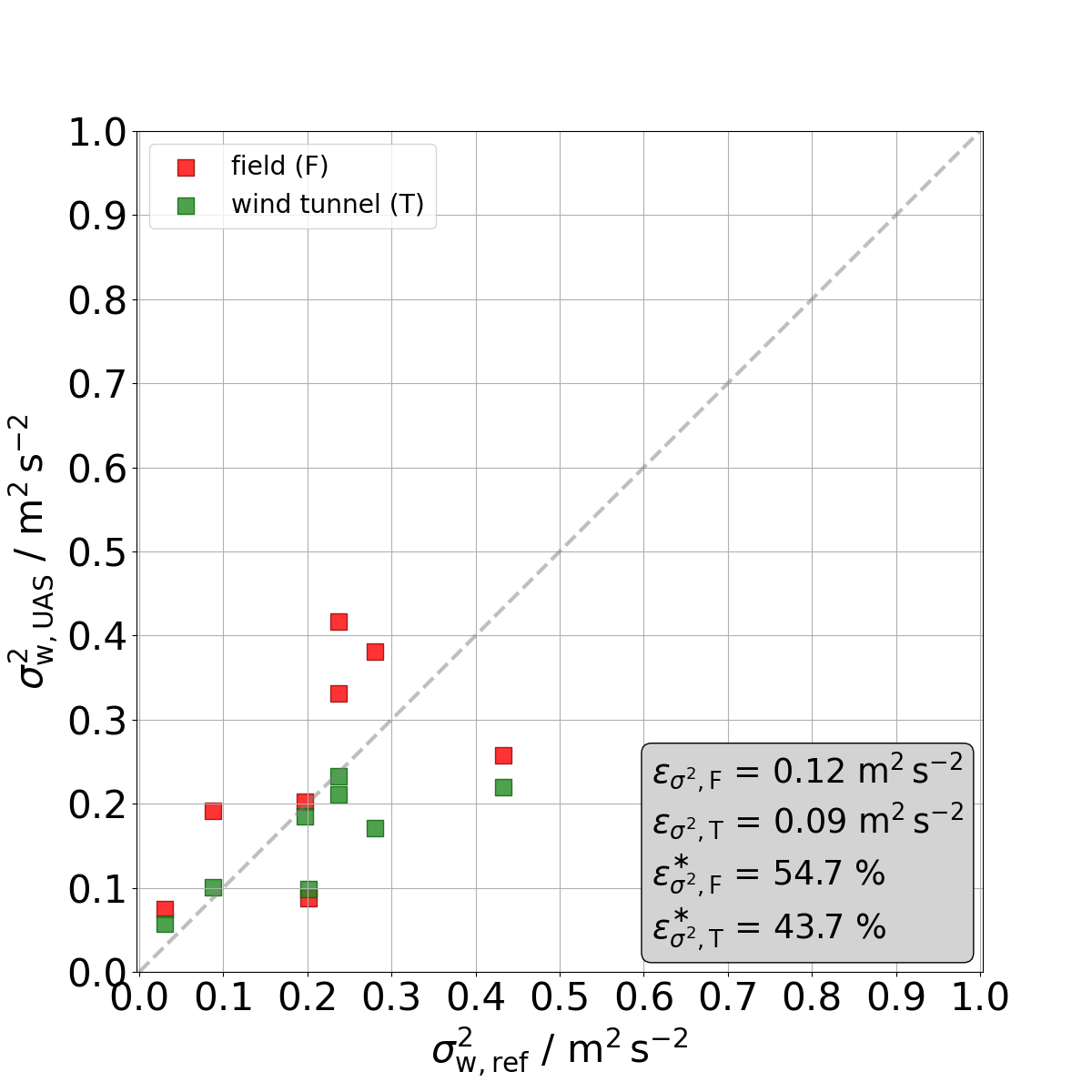}
    \caption{Comparison of the variance of the vertical wind $\sigma^2_{\mathrm{w}}$ determined using the calibration coefficients from the field calibration (red) by \cite{Wildmann2022} and the wind tunnel calibration (green) for the 5.0$^{\prime\prime}$ setup.}
 \label{fig:scatter_field_vs_lab}
  \end{subfigure}\hfill
  \begin{subfigure}{0.48\textwidth}
    \centering
    \includegraphics[width=\linewidth]{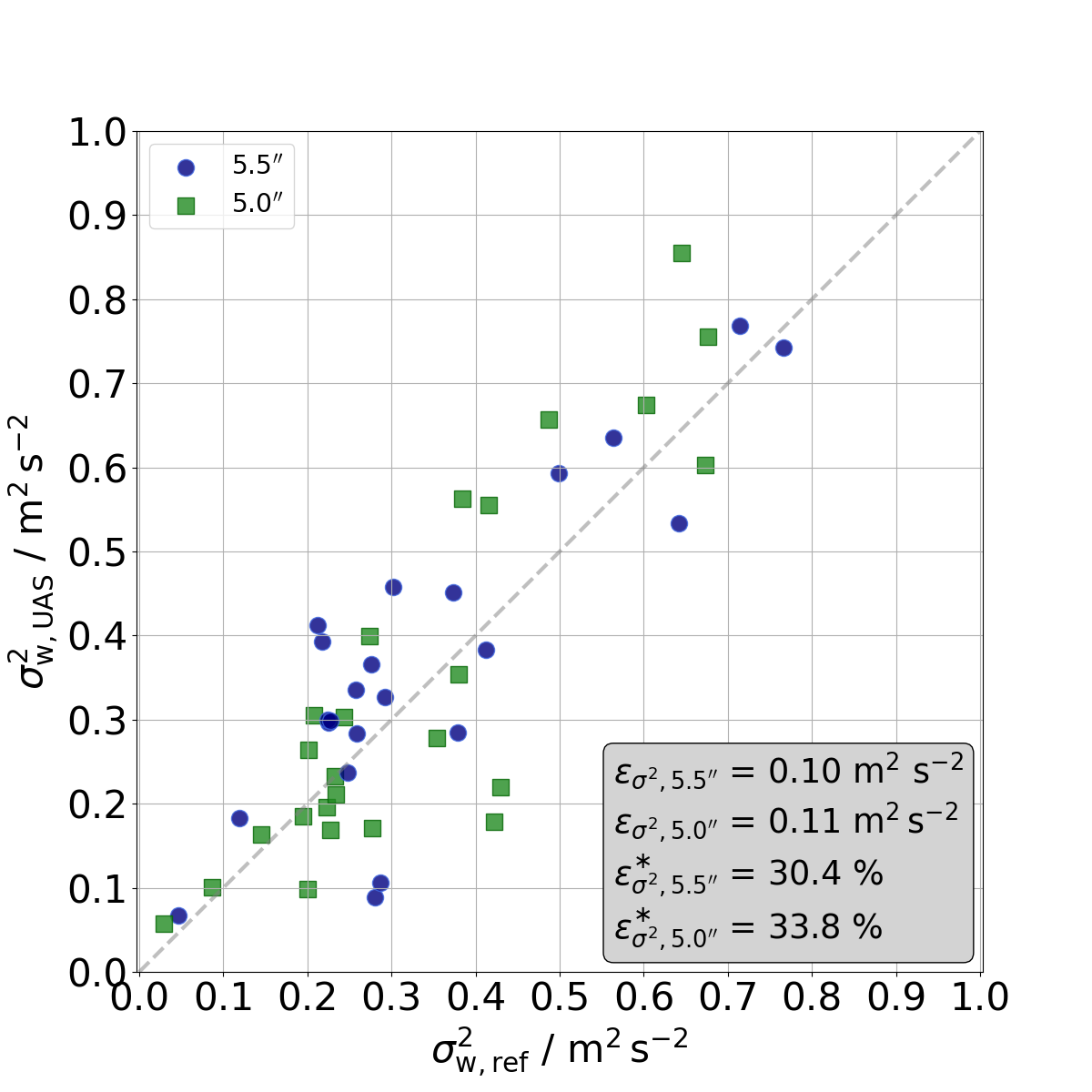}
    \caption{Comparison of the variance of the vertical wind $\sigma^2_{\mathrm{w}}$ determined using with UASs in the 5.0$^{\prime\prime}$ setup (green) and the 5.5$^{\prime\prime}$ setup (blue).\newline}
  \label{fig:scatter_5_vs_5p5}
  \end{subfigure}
  \caption{}
\end{figure}

The comparison of the speed curves between the 5.5$^{\prime\prime}$ and 5.0$^{\prime\prime}$ propeller setups, as well as the wind tunnel calibration and field calibration by \cite{Wildmann2022} of the 5.0$^{\prime\prime}$ propeller setup, is shown in Fig.~\ref{fig:cal_curve_w_comparison}. With this calibration, the full 3D wind vector can now be determined based on the avionics data (see Eq. \ref{eq:gesamt}-\ref{eq:1.3c}).  
\newline
The comparison of the variance estimation of the vertical wind component with the calibration coefficients from the wind tunnel and from the field measurements by \cite{Wildmann2022}, applied to the avionics data for measurements of the setup comparison flights with $\leq 8~\mathrm{m~s^{-1}}$ horizontal wind, is shown in Fig.~\ref{fig:scatter_field_vs_lab}. Due to the speed limit, only 8 flights with this setup are available for comparison. There are no wind speed limitations on the application of the calibration coefficients from the wind tunnel, which is why they can be used for all measurements of the two UAS setups from the setup comparison flights. The variance determination of the $w$ component  with the different setups against reference measurements  is shown in Fig.~\ref{fig:scatter_5_vs_5p5}.  

\begin{figure}[H]
  \centering
  \begin{subfigure}{0.432\textwidth}
    \centering
    \includegraphics[width=\linewidth]{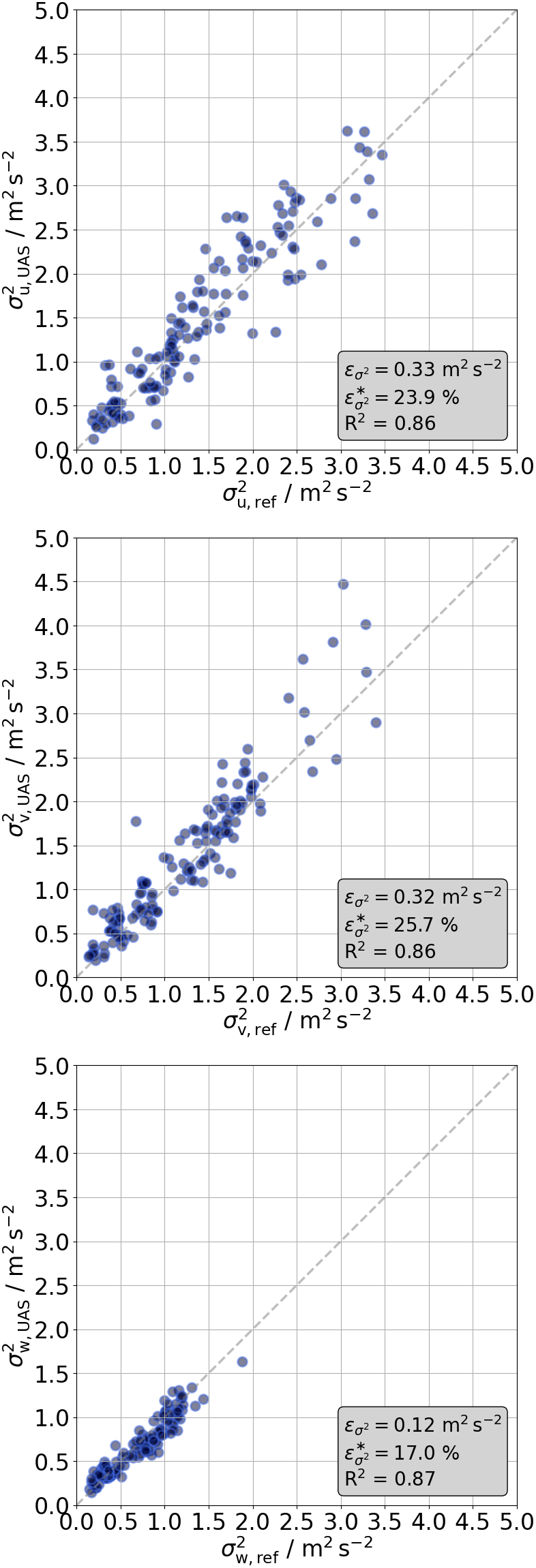}
    \caption{Variances $\sigma^2$ }
    \label{fig:scatter_var_components}
  \end{subfigure}
  \begin{subfigure}{0.4455\textwidth}
    \centering
    \includegraphics[width=\linewidth]{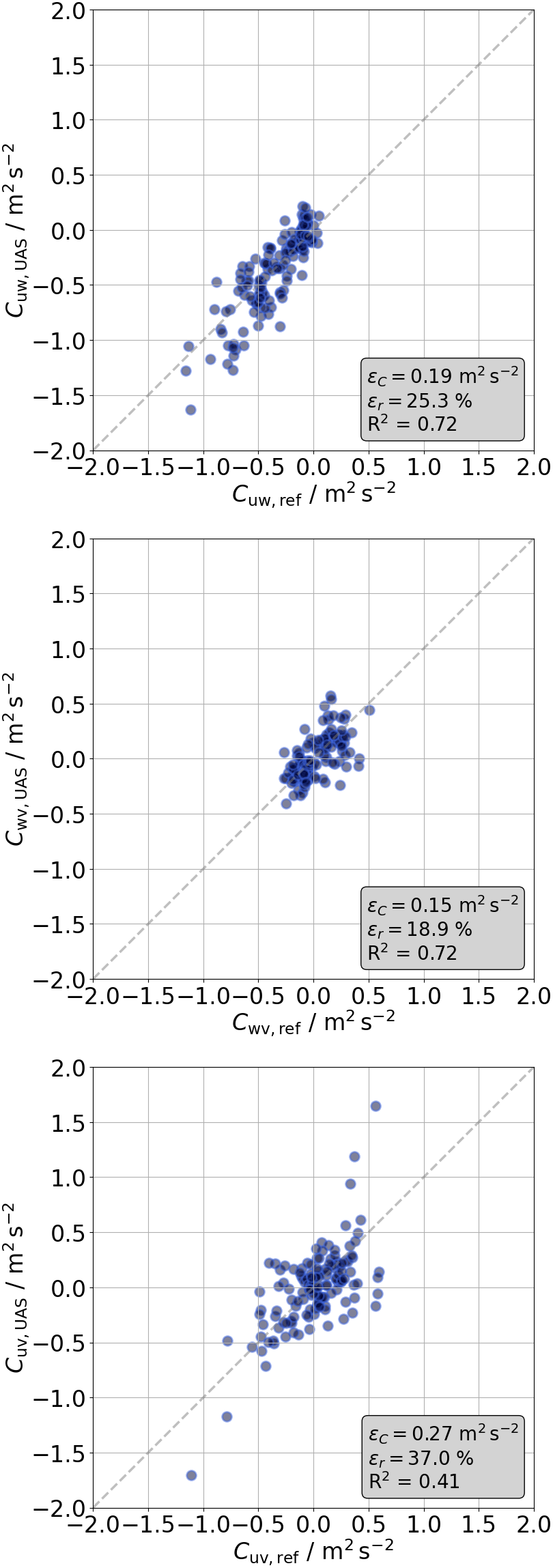}
    \caption{Covariances $C$ }
  \label{fig:scatter_covar_components}
  \end{subfigure}
  \caption{UAS measurements versus the respective sonic anemometer for the individual 3D wind speed components.}
  
\end{figure}

The RMSE for the variance measurements of $w$ with the 5.5$^{\prime\prime}$ setup from the box-shape flights confirms the RMSE of $\sim$0.1~$\mathrm{m^2\,s^{-2}}$ from the setup comparison flights and shows a lower nRMSE with almost three times as many data points (see Fig.~\ref{fig:scatter_var_components}). The RMSE for mean $w$ in these flights is 0.18~$\mathrm{m\,s^{-1}}$. 
\newpage

\begin{figure}[H]
  \centering
  \begin{subfigure}{0.48\textwidth}
    \centering
    \includegraphics[width=\linewidth]{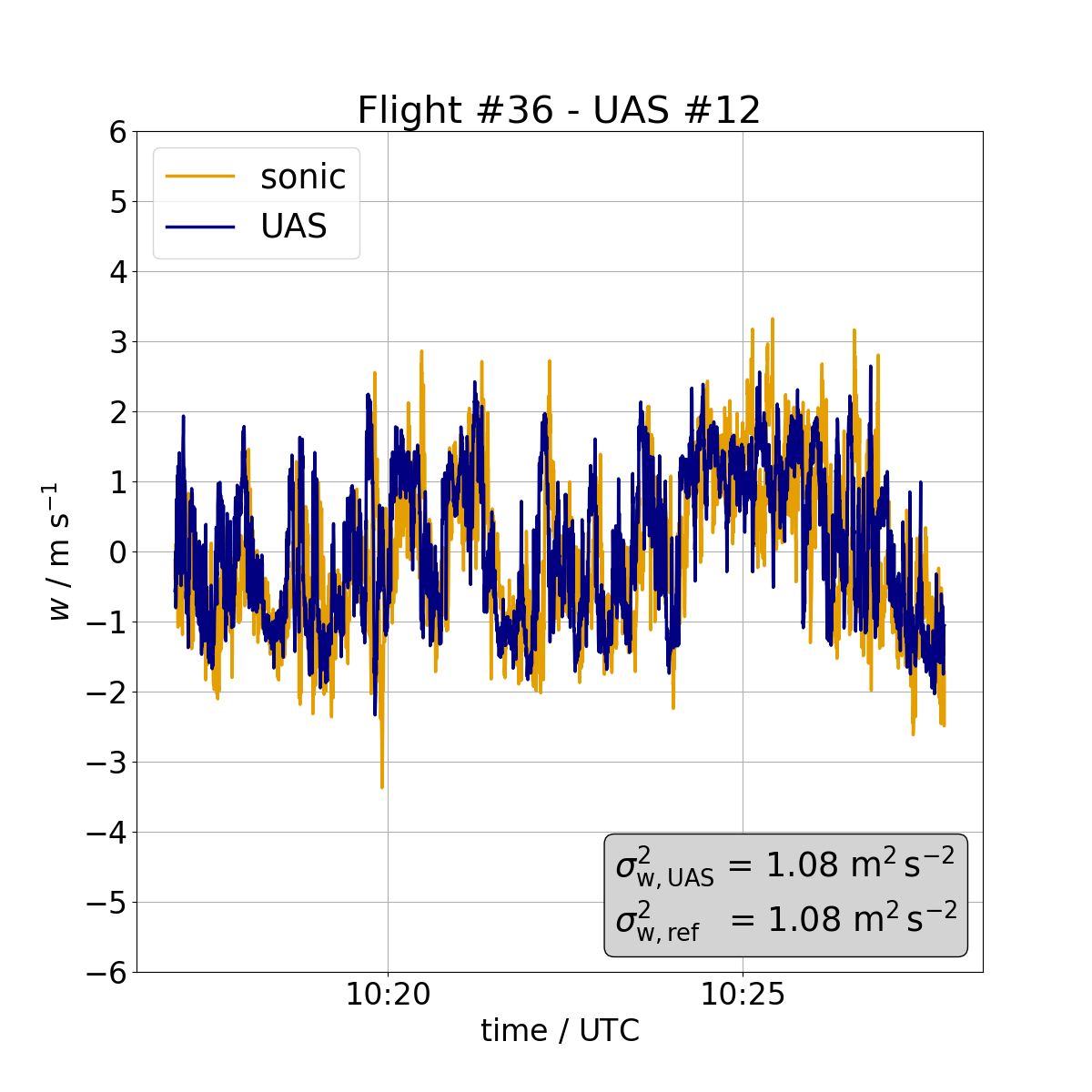}
    \caption{Time series of vertical wind speed $w$. \newline \newline \newline  \newline}
    \label{fig:ts_field_comparison}
  \end{subfigure}\hfill
  \begin{subfigure}{0.48\textwidth}
    \centering
    \includegraphics[width=\linewidth]{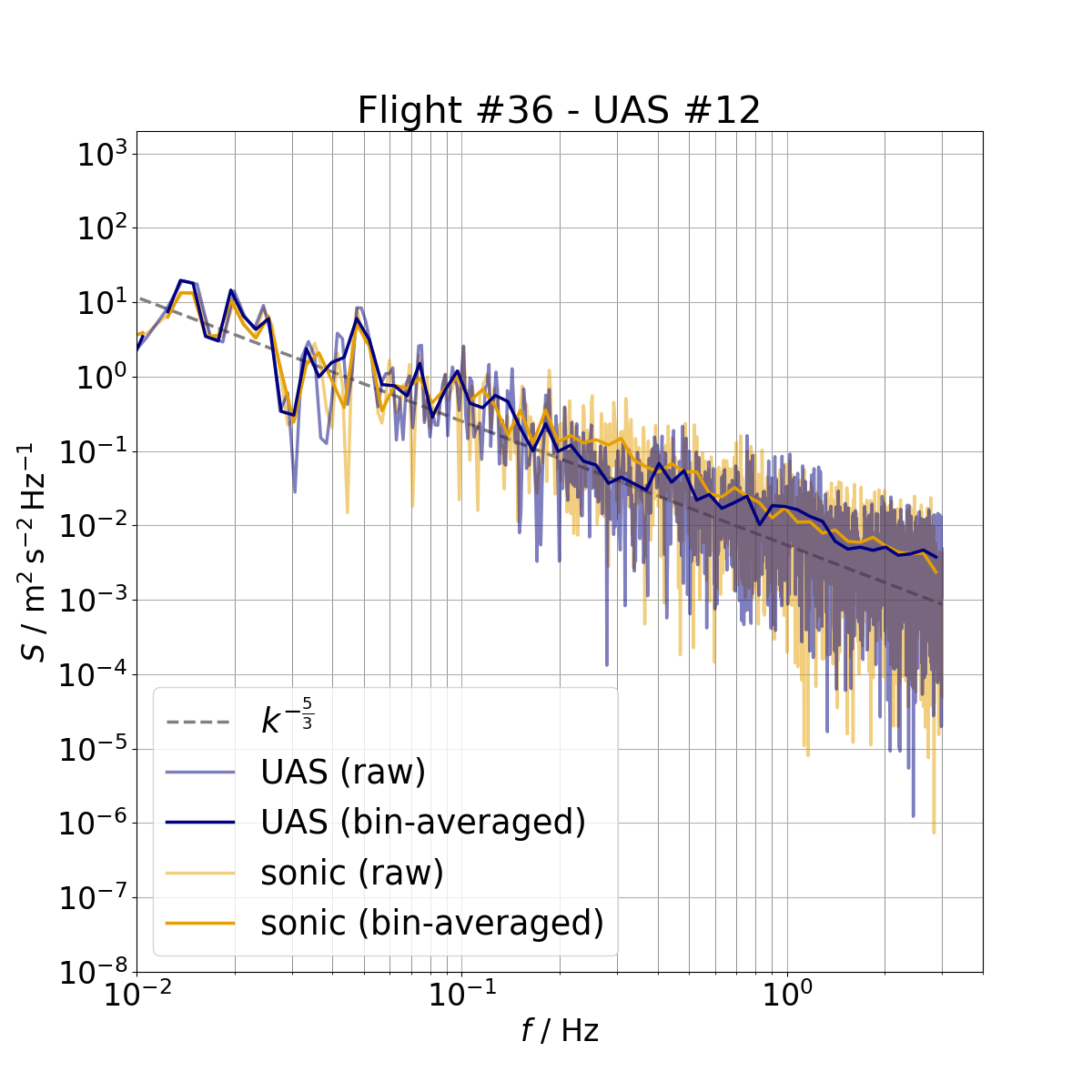}
    \caption{PSD $S$ over the frequency $f$; the transparent curves show the raw spectra, the opaque curves overlaid on them show the bin-averaged spectra using 85 logarithmically spaced frequency bins. The dashed gray line represents the spectrum following the Kolmogorov $-\,\nicefrac{5}{3}$ power law.}
  \label{fig:w_spectra}
  \end{subfigure}
  \caption{Exemplary comparison of the vertical wind measured via UAS (blue) versus sonic anemometer (orange).}
  \label{fig:field_comparison_ts_psd}
\end{figure}

Furthermore, the variances of the 3D wind components and their covariance behavior with respect to each other have now been determined using UAS measurements, and the measurement uncertainty has been quantified in comparison to the reference sensors (see Fig.~\ref{fig:scatter_covar_components}).
A more detailed investigation of the turbulence is possible by determining the PSD of the $w$ component with the drone. A comparison of the time series and its PSD from the UAS measurements with the reference measurements of the sonic anemometers at the measurement mast array are shown in Fig.~\ref{fig:field_comparison_ts_psd}.

\begin{figure}[H]
  \centering
  \includegraphics[width=0.48\textwidth]{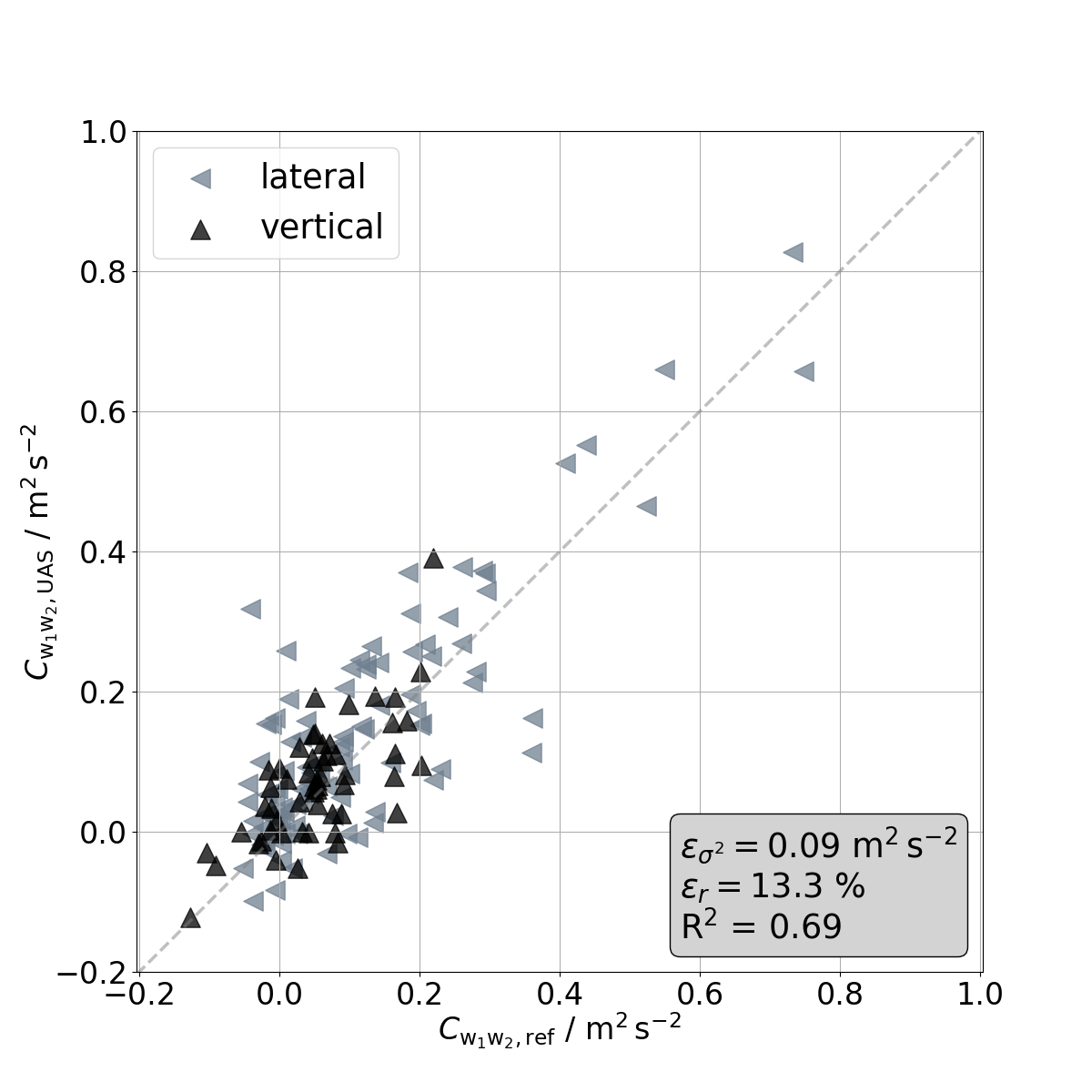} 
  \caption{Covariances of the vertical wind component $C_{\mathrm{w_1,w_2}}$ measured by laterally (gray) or vertically (black) adjacent drones versus the corresponding sonic anemometers.}
  \label{fig:scatter_cov_latalt}
\end{figure}

The PSD is a key factor in the covariance of different measurement points in fleet measurements. The covariances of the vertical wind components measured by drones adjacent to the main wind direction, both vertically and laterally, are shown in Fig.~\ref{fig:scatter_cov_latalt}.

  \section{Discussion}
In the results of the wind tunnel measurements it can be seen from $a_{\mathrm{L1}}$ and $a_{\mathrm{L2}}$ in Table \ref{tab:hor_coeff} that the change in flight behavior is very similar for both propeller setups with regard to the horizontal wind.
The differences in the degree of polynomials used to model lift behavior over horizontal wind in the lateral and longitudinal direction are particularly noticeable in the lower and mid speed range, while the behavior of the two curves is relatively similar in higher ranges (see Fig.~\ref{fig:lift_curve_comparison_x_y}). 
We attribute this to downstream effects of the fuselage at low speeds and thus at low tilt angles: While all four motors are exposed to free flow during frontal flow and therefore generate aerodynamic lift, two propellers are in the slipstream of the frame under lateral flow. These effects get smaller with increasing tilt angles as the propellers move out of the fuselage's blocking. 
\newline
In the range where the horizontal airflow of the drone generates lift, the lift curves between the 5.0$^{\prime\prime}$ and 5.5$^{\prime\prime}$ setups initially differ very little. With both setups, lift is maximum at similar longitudinal wind forces; only beyond this point are the vertical forces significantly higher with the 5.5$^{\prime\prime}$ setup (initially as lift, then as downforce).
\newline
The free-field measurements with the 5.0$^{\prime\prime}$ and 5.5$^{\prime\prime}$ setups show similar measurement behavior and accuracies (see Fig.~\ref{fig:scatter_5_vs_5p5}), which indicates the robustness and transferability of our calibration method. 
Consistent accuracies between the two measurement campaigns (Fig.~\ref{fig:scatter_5_vs_5p5} and Fig.~\ref{fig:scatter_covar_components}) with the 5.5$^{\prime\prime}$ setup demonstrate the reproducibility and validity of the calibration, answering research question 1. 
\newline
With an $0.18~\mathrm{m^2~s^{-2}}$ for measuring vertical wind, an RMSE of $0.12~\mathrm{m^2~s^{-2}}$ and an nRMSE of $17.0~\%$ for determining its variance, and an RMSE of  $<0.1~\mathrm{m^2\,s^{-2}}$ for covariances between different measurement points, our calibration method achieves a high level of accuracy in field measurements. With regard to research question 2, this shows that the wind tunnel calibration can be transferred to fleet measurements. 
\newline
These accuracies for the vertical wind as well as RMSEs of $\sim$0.3~$\mathrm{m^2\,s^{-2}}$ and nRMSEs of $\sim$25~\%  for the variance of the horizontal wind measurement were found, and RMSEs below 0.3~$\mathrm{m^2\,s^{-2}}$ for the covariances of all wind components determined, thereby achieving satisfactory uncertainties in the 3D wind measurements and answering research question 3. The PSDs of the vertical wind component measured with drones agree well with the PSDs measured with reference sensors up to high frequency ranges.  
\newline
Nevertheless, there is a certain degree of uncertainty in our wind tunnel calibration method: The vertical gusts generated develop along the longitudinal movement along the wind tunnel. While the reference was always placed firmly in one position, the drone was not able to maintain a constant distance from the grid in every situation and recorded the gusts in different stages of development in the longitudinal movement. The resulting uncertainty contributes to the scatter of the data points for the curve fit from $w$ to $F_{\mathrm{z}}$ shown in Fig.~\ref{fig:cal_curve_w}. 
\newline
In field measurements, various factors introduce uncertainties into the measurement setup, which contribute to the RMSEs of the measurements. Firstly, the distance between UAS and reference sensor is 50~m, and depending on the wind direction, the sensors are not perfectly aligned along the wind direction. However, we cannot separate this error value from the quantification of the uncertainty of our turbulence measurement with the drone.
\newline
Furthermore, after taking the measurements, we discovered that the selected reference sensor on the center mast at 25~m was faulty, which is why we had to use the next lowest sensor on the mast (i.e., at a height of 20~m) as a reference for the corresponding UASs, which can increase uncertainty due to differences in measurement height. 
\newline
One source of uncertainty in our sensor technology is the error in the drone's yaw angle. If the magnetometer provides incorrect orientation data to the autopilot, this is incorporated into the equations of our wind algorithm via the avionics data, resulting in incorrect calculations of the components' contributions to the total wind vector. This is reflected in the covariance analyses of the wind components; the RMSEs of the UAS measurements compared to the reference sensors is highest for $C_{\mathrm{uv}}$.
\newline
Our results differ from those of \cite{Wildmann2022} in several respects: Firstly, the lift curves differ significantly (see Fig.~\ref{fig:lift_curve_comparison}). This is partly due to the fact that, in contrast to the field conditions, the defined wind conditions in the wind tunnel are $w_{\mathrm{g}}=0$.
It is also due to the lack of data for high wind speeds in the calibrations by \cite{Wildmann2022}, whereas we were able to generate both low and high wind speeds in the wind tunnel. 
\newline
These advances due to wind tunnel calibration lead to improvements  of the vertical wind measurement in comparison to previous field calibration, answering research question 4:
As a first impact, we can work without a threshold value for horizontal wind speed above which vertical wind measurement is no longer applicable. Accordingly, the measured vertical wind speed is defined for the entire measurement period, which is why, unlike \cite{Kistner2024} and \cite{Wildmann2022}, we can determine the complete 3D vector for all wind speeds we operate the drones in. 
\newline
The different lift coefficients also have an effect on the calibration curve for wind speed. While ours and those of \cite{Wildmann2022} are similar in the lower speed ranges, the deviations in the higher speed ranges are very distinct (see Fig.~\ref{fig:cal_curve_w_comparison}). This is also due to the fact that more data is available for fitting at high wind speeds for this calibration in the wind tunnel than for field calibrations. Measuring larger vertical wind speed components is important when it comes to determining vertical exchange. This requires accurate measurements even at higher vertical wind speeds, which is where our calibration makes an important contribution.
Furthermore, the wind tunnel calibrations enabled us to determine the lift generated by the wind component that hits the UAS laterally. This leads to a more precise determination of vertical wind in crosswinds. 
\newline
Our RMSE for vertical wind measurement of 0.18~$\mathrm{m\,s^{-1}}$ is higher as that of \cite{Wildmann2022} at 0.14~$\mathrm{m\,s^{-1}}$, but with more extreme vertical wind speeds. It lies within the range of 0.15 to 0.4~$\mathrm{m\,s^{-1}}$ that \cite{Thielicke2021} achieved and is below the value of 0.49~$\mathrm{m\,s^{-1}}$ reported by \cite{Segales2025}. A comparison with \cite{Zhu2025} is not possible in this respect, as they only compare their measurements with their reference sonic anemometer for horizontal wind and thus provide an RMSE; they do not give this value for vertical wind measurement. 
\newline
In conclusion, we could confirm our hypothesis in field validations that wind tunnel calibration enables us to overcome the limitations in vertical wind measurement. 

\section{Conclusion}
In this work, we successfully performed advanced calibration of vertical wind measurement using UASs under laboratory conditions for two UAS setups in a wind tunnel with active grid and validated these in field measurements. For the first time, a multicopter drone was flown systematically in wind tunnel gusts with varying vertical component. A unique experimental setup with special programs for the active grid was developed for this purpose. For our small drones, this allows a full-scale experiment in a controlled environment. The full operational wind speed range for the drone could be covered in the wind tunnel. Direct spatial comparisons of drone measurements with in situ instrumentation could be enabled in the WiValdi research wind farm with the unique mast array.
To perform the wind tunnel calibrations, we first determined the calibration coefficients for horizontal wind determination in the drone's fixed coordinate system. We were then able to determine lift as a function of horizontal wind from wind speeds up to $18~\mathrm{m~s^{-1}}$ for the longitudinal and lateral flow around the drone in the absence of vertical wind. This allowed us to overcome the previous limitation of $8~\mathrm{m~s^{-1}}$ as the maximum speed for the application of vertical wind measurement. 
By taking measurements in these vertical gusts generated by the active grid, we were able to determine the calibration coefficients for vertical wind determination at different vertical wind speeds with a wide distribution of approximately $-3.5\dots5.5~\mathrm{m~s^{-1}}$ at various horizontal winds up to $17.5~\mathrm{m~s^{-1}}$. 
As a result of these wind tunnel calibrations, we were able to enable the measurement of the complete 3D wind vector for all operational wind speeds using deterministic methods with the drone as a sensor.
\newline
We also validated this measurement method in fleet flights with up to 10 drones simultaneously against sonic anemometers. When determining the variance of the vertical wind at flights within the limitations of field calibration, we were able to achieve a comparable accuracy of wind tunnel calibration as with field calibration.  Comparative measurements between different UAS setups without limitations demonstrate that our calibration method is generalizable for different hardware. In fleet flights with a chosen setup, it was possible to show with significantly more data points that our calibration is stable and reproducible, and can be transferred to fleet measurements. In general, high accuracies are achieved in the determination of turbulence parameters such as variance and covariance of the individual 3D wind components, especially the vertical wind. The PSD of the vertical wind measured in high resolution with the drones agrees well with the reference measurements, which is also reflected in high covariances of our multi-point measurements with the fleet.
\newline
Accurate vertical wind measurements are important to measure complex flow structures and turbulent transport in the ABL. The SWUF-3D fleet has already been deployed in campaigns studying wake dynamics behind wind turbines and valley flows in Alpine valleys. The possibilities for future analysis of the measurement data from these campaigns are improved by the contribution of this work as accurate estimation of the vertical flow velocity is particularly important in those conditions with highly 3D flow.  
In summary, with our wind tunnel calibrations it was feasible to overcome previous limitations and to enable precise, high-resolution 3D in situ wind measurements using a fleet of drones.

\clearpage
\acknowledgments
This research is part of the project ESTABLIS-UAS and has been supported by the HORIZON EUROPE European Research Council (grant no. 101040823). 
Many thanks to Almut Alexa and Thomas Messmer for their intensive assistance and support during the wind tunnel experiments and to Lars Neuhaus for providing the protocols used to control the active grid. We thank Almut Alexa and Laszlo Györy for their efforts in supporting the field measurements. Paul Waldmann reviewed the manuscript internally and we thank him for his valuable comments.
We also thank Michael Hölling for his help in preparing the wind tunnel measurements and Joachim Peinke for his helpful advice to this work. 
Pixhawk is a trademark of Lorenz Meier. 


%
%
\datastatement

The data is available under doi.org/10.5281/zenodo.17700905.

%

\appendix



\appendix[A] 
\appendixtitle{Nomenclature}
\begin{longtable}{ll}
\endfirsthead
            $a$ & acceleration \\
            ABL & atmospheric boundary layer \\
            $b$ & exponential calibration coefficient \\
            $C$ & covariance \\
            $c$ &  linear calibration coefficient \\
            $f$ & frequency \\
            $F_{\textrm{L}}$ & lift force \\
            $F_{\textrm{T}}$ & thrust force \\
            $F_{\textrm{z}}$ & vertical force \\
            GNSS & global navigation satellite system\\
            $m$ & mass\\
            nRMSE & normalized root mean square error\\
            PSD & power spectral density \\
            $\textbf{\textit{R}}$& rotation matrix \\
            $R^2$ & coefficient of determination \\
            $r$ & Pearson correlation coefficient \\
            RMSE & root mean square error \\
            $S$  & PSD\\       
            $s$ & exemplary for any but defined variable \\
            $t$ & time \\
            $u$ & longitudinal wind speed component \\
            UAS & uncrewed aerial system \\
            $\textbf{\textit{V}}$ & wind speed vector \\
            $V$ & wind speed \\
            $v$ & lateral wind speed component \\
            $w$ & vertical wind speed component \\
            $\varepsilon$ &  RMSE \\
            $\varepsilon^{\ast}$ &  nRMSE \\
            $\theta$ & pitch angle \\
            $\sigma^2$ & variance\\
            $\tau$ & time increment \\
            $\phi$ & roll angle  \\
            $\psi$ & yaw angle  \\
            $\omega$ & motor rotational speed\\
\end{longtable}

%



\bibliographystyle{ametsocV6}
\bibliography{references}

@Article{Palomaki2017,
  author    = {Ross T. Palomaki and Nathan T. Rose and Michael van den Bossche and Thomas J. Sherman and Stephan F. J. De Wekker},
  journal   = {Journal of Atmospheric and Oceanic Technology},
  title     = {Wind Estimation in the Lower Atmosphere Using Multirotor Aircraft},
  year      = {2017},
  month     = {may},
  number    = {5},
  pages     = {1183--1191},
  volume    = {34},
  doi       = {10.1175/jtech-d-16-0177.1},
  publisher = {American Meteorological Society},
}

@Article{GonzalezRocha2019,
  author     = {Javier Gonz{\'{a}}lez-Rocha and Craig A. Woolsey and Cornel Sultan and Stephan F. J. De Wekker},
  journal    = {Journal of Guidance, Control, and Dynamics},
  title      = {Sensing Wind from Quadrotor Motion},
  year       = {2019},
  month      = {apr},
  number     = {4},
  pages      = {836--852},
  volume     = {42},
  doi        = {10.2514/1.g003542},
  publisher  = {American Institute of Aeronautics and Astronautics ({AIAA})},
  readstatus = {skimmed},
}

@Article{Thielicke2021,
  author     = {William Thielicke and Waldemar Hübert and Ulrich Müller and Michael Eggert and Paul Wilhelm},
  journal    = {Atmospheric Measurement Techniques},
  title      = {Towards accurate and practical drone-based wind measurements with an ultrasonic anemometer},
  year       = {2021},
  month      = {feb},
  number     = {2},
  pages      = {1303--1318},
  volume     = {14},
  doi        = {10.5194/amt-14-1303-2021},
  publisher  = {Copernicus {GmbH}},
  readstatus = {read},
}

@Article{Neumann2015,
  author    = {Patrick P. Neumann and Matthias Bartholmai},
  journal   = {Sensors and Actuators A: Physical},
  title     = {Real-time wind estimation on a micro unmanned aerial vehicle using its inertial measurement unit},
  year      = {2015},
  month     = {nov},
  pages     = {300--310},
  volume    = {235},
  doi       = {10.1016/j.sna.2015.09.036},
  publisher = {Elsevier {BV}},
}

@Article{Brosy2017,
  author    = {Caroline Brosy and Karina Krampf and Matthias Zeeman and Benjamin Wolf and Wolfgang Junkermann and Klaus Schäfer and Stefan Emeis and Harald Kunstmann},
  title     = {Simultaneous multicopter-based air sampling and sensing of meteorological variables},
  doi       = {10.5194/amt-10-2773-2017},
  number    = {8},
  pages     = {2773--2784},
  volume    = {10},
  journal   = {Atmospheric Measurement Techniques},
  month     = {aug},
  publisher = {Copernicus {GmbH}},
  year      = {2017},
}

@Article{Wildmann2015,
  author    = {Norman Wildmann and Gerrit Anke Rau and Jens Bange},
  journal   = {Boundary-Layer Meteorology},
  title     = {Observations of the Early Morning Boundary-Layer Transition with Small Remotely-Piloted Aircraft},
  year      = {2015},
  month     = {aug},
  number    = {3},
  pages     = {345--373},
  volume    = {157},
  doi       = {10.1007/s10546-015-0059-z},
  publisher = {Springer Science and Business Media {LLC}},
}

@Article{Wetz2021,
  author     = {Tamino Wetz and Norman Wildmann and Frank Beyrich},
  journal    = {Atmospheric Measurement Techniques},
  title      = {Distributed wind measurements with multiple quadrotor unmanned aerial vehicles in the atmospheric boundary layer},
  year       = {2021},
  month      = {may},
  number     = {5},
  pages      = {3795--3814},
  volume     = {14},
  doi        = {10.5194/amt-14-3795-2021},
  publisher  = {Copernicus {GmbH}},
  readstatus = {read},
}

@Article{Wetz2022,
  author     = {Tamino Wetz and Norman Wildmann},
  title      = {Spatially distributed and simultaneous wind measurements with a fleet of small quadrotor {UAS}},
  doi        = {10.1088/1742-6596/2265/2/022086},
  number     = {2},
  pages      = {022086},
  volume     = {2265},
  journal    = {Journal of Physics: Conference Series},
  month      = {may},
  publisher  = {{IOP} Publishing},
  readstatus = {read},
  year       = {2022},
}

@Article{Neuhaus2021,
  author     = {Lars Neuhaus and Frederik Berger and Joachim Peinke and Michael Hölling},
  journal    = {Experiments in Fluids},
  title      = {Exploring the capabilities of active grids},
  year       = {2021},
  month      = {may},
  number     = {6},
  volume     = {62},
  doi        = {10.1007/s00348-021-03224-5},
  publisher  = {Springer Science and Business Media {LLC}},
  readstatus = {skimmed},
}

@Article{Segales2020,
  author    = {Antonio R. Segales and Brian R. Greene and Tyler M. Bell and William Doyle and Joshua J. Martin and Elizabeth A. Pillar-Little and Phillip B. Chilson},
  journal   = {Atmospheric Measurement Techniques},
  title     = {The {CopterSonde}: an insight into the development of a smart unmanned aircraft system for atmospheric boundary layer research},
  year      = {2020},
  month     = {may},
  number    = {5},
  pages     = {2833--2848},
  volume    = {13},
  doi       = {10.5194/amt-13-2833-2020},
  publisher = {Copernicus {GmbH}},
}

@Article{Rautenberg2019,
  author    = {Alexander Rautenberg and Jonas Allgeier and Saskia Jung and Jens Bange},
  journal   = {Atmosphere},
  title     = {Calibration Procedure and Accuracy of Wind and Turbulence Measurements with Five-Hole Probes on Fixed-Wing Unmanned Aircraft in the Atmospheric Boundary Layer and Wind Turbine Wakes},
  year      = {2019},
  month     = {mar},
  number    = {3},
  pages     = {124},
  volume    = {10},
  doi       = {10.3390/atmos10030124},
  publisher = {{MDPI} {AG}},
}

@Article{GonzalezRocha2020,
  author    = {Javier Gonz{\'{a}}lez-Rocha and Stephan F. J. De Wekker and Shane D. Ross and Craig A. Woolsey},
  journal   = {Sensors},
  title     = {Wind Profiling in the Lower Atmosphere from Wind-Induced Perturbations to Multirotor {UAS}},
  year      = {2020},
  month     = {feb},
  number    = {5},
  pages     = {1341},
  volume    = {20},
  doi       = {10.3390/s20051341},
  publisher = {{MDPI} {AG}},
}

@Article{Hattenberger2022,
  author    = {Gautier Hattenberger and Murat Bronz and Jean-Philippe Condomines},
  title     = {Estimating wind using a quadrotor},
  doi       = {10.1177/17568293211070824},
  pages     = {175682932110708},
  volume    = {14},
  journal   = {International Journal of Micro Air Vehicles},
  month     = {jan},
  publisher = {{SAGE} Publications},
  year      = {2022},
}

@Article{Marino2015,
  author    = {Matthew Marino and Alex Fisher and Reece Clothier and Simon Watkins and Samuel Prudden and Chung Sing Leung},
  journal   = {International Journal of Micro Air Vehicles},
  title     = {An Evaluation of Multi-Rotor Unmanned Aircraft as Flying Wind Sensors},
  year      = {2015},
  month     = {sep},
  number    = {3},
  pages     = {285--299},
  volume    = {7},
  doi       = {10.1260/1756-8293.7.3.285},
  publisher = {{SAGE} Publications},
}

@Article{Kistner2024,
  author       = {Kistner, Johannes and Neuhaus, Lars and Wildmann, Norman},
  date         = {2024-08},
  journal = {Atmospheric Measurement Techniques},
  title        = {High-resolution wind speed measurements with quadcopter uncrewed aerial systems: calibration and verification in a wind tunnel with an active grid},
  doi          = {10.5194/amt-17-4941-2024},
  issn         = {1867-8548},
  number       = {16},
  pages        = {4941--4955},
  volume       = {17},
  publisher    = {Copernicus GmbH},
  year      = {2024},
}

@Article{Wildmann2024,
  author       = {Wildmann, N and Kistner, J},
  date         = {2024-06},
  journal = {Journal of Physics: Conference Series},
  title        = {An evaluation of different measurement strategies to measure wind turbine near wake flow with small multicopter UAS},
  doi          = {10.1088/1742-6596/2767/4/042004},
  issn         = {1742-6596},
  number       = {4},
  pages        = {042004},
  volume       = {2767},
  publisher    = {IOP Publishing},
  year      = {2024},
}

@Article{Wildmann2025,
  author       = {Wildmann, Norman and Kistner, Johannes},
  date         = {2025-05},
  journal = {Journal of Physics: Conference Series},
  title        = {In situ measurements of near wake dynamics with a fleet of multicopter drones},
  doi          = {10.1088/1742-6596/3016/1/012011},
  issn         = {1742-6596},
  number       = {1},
  pages        = {012011},
  volume       = {3016},
  year      = {2025},
  publisher    = {IOP Publishing},

}

@Article{Zhu2025,
  author       = {Zhu, Sihong and Zhao, Tonghui and Zhang, Huanji and Chen, Yichao and Yang, Dongxu and Liu, Yi and Cao, Junji},
  date         = {2025-06},
  journal = {Drones},
  title        = {UAVs’ Flight Dynamics Is All You Need for Wind Speed and Direction Measurement in Air},
  doi          = {10.3390/drones9070466},
  issn         = {2504-446X},
  number       = {7},
  pages        = {466},
  volume       = {9},
  year      = {2025},
  publisher    = {MDPI AG},
}

@Article{Yang2025,
  author       = {Yang, Yanrong and Zhang, Yuheng and Han, Tianran and Xie, Conghui and Liu, Yayong and Huang, Yufei and Zhou, Jietao and Sun, Haijiong and Zhao, Delong and Zhang, Kui and Li, Shao-Meng},
  date         = {2025-07},
  journal = {Atmospheric Measurement Techniques},
  title        = {A correction algorithm for rotor-induced airflow and flight attitude changes during three-dimensional wind speed measurements made from a rotary unoccupied aerial vehicle},
  doi          = {10.5194/amt-18-3035-2025},
  issn         = {1867-8548},
  number       = {13},
  pages        = {3035--3050},
  volume       = {18},
  publisher    = {Copernicus GmbH},
  year      = {2025},
}

@Article{Wildmann2025a,
  author       = {Wildmann, Norman and Györy, Laszlo},
  date         = {2025-10},
  journal = {Atmospheric Measurement Techniques},
  title        = {Towards sensible heat flux measurements with fast-response fine-wire platinum resistance thermometers on small multicopter uncrewed aerial systems},
  doi          = {10.5194/amt-18-5527-2025},
  issn         = {1867-8548},
  number       = {20},
  pages        = {5527--5544},
  volume       = {18},
  publisher    = {Copernicus GmbH},
  year      = {2025},
}

@Article{Wildmann2022,
  author       = {Wildmann, Norman and Wetz, Tamino},
  date         = {2022-09},
  journal = {Atmospheric Measurement Techniques},
  title        = {Towards vertical wind and turbulent flux estimation with multicopter uncrewed aircraft systems},
  doi          = {10.5194/amt-15-5465-2022},
  issn         = {1867-8548},
  number       = {18},
  pages        = {5465--5477},
  volume       = {15},
  publisher    = {Copernicus GmbH},
  year      = {2022},
}

@Article{Ghirardelli2025,
  author       = {Ghirardelli, Mauro and Kral, Stephan T. and Cheynet, Etienne and Reuder, Joachim},
  date         = {2025-05},
  journal = {Atmospheric Measurement Techniques},
  title        = {SAMURAI-S: Sonic Anemometer on a MUlti-Rotor drone for Atmospheric turbulence Investigation in a Sling load configuration},
  doi          = {10.5194/amt-18-2103-2025},
  issn         = {1867-8548},
  number       = {9},
  pages        = {2103--2124},
  volume       = {18},
  publisher    = {Copernicus GmbH},
  year      = {2025},
}

@Article{Bramati2025,
  author       = {Bramati, Matteo and Savvakis, Vasileios and Beyrich, Frank and Bange, Jens and Platis, Andreas},
  date         = {2025-03},
  journal = {Journal of Atmospheric and Oceanic Technology},
  title        = {In Situ Uncrewed Aircraft Measurements of Turbulent Kinetic Energy over Heterogeneous Terrain},
  doi          = {10.1175/jtech-d-24-0026.1},
  issn         = {1520-0426},
  number       = {3},
  pages        = {319--338},
  volume       = {42},
  year      = {2025},
  publisher    = {American Meteorological Society},
}

@thesis{Hofmann2025,
            year = {2025},
          author = {Hofmann, Paula},
           month = {Februar},
           title = {Weiterentwicklung der Beobachtungsf{\"a}higkeit von Methanemissionen {\"u}ber drohnengest{\"u}tzte Messungen mittes low-cost Sensoren},
            type = {Diplomarbeit},
          school = {Technische Universit{\"a}t Dresden},
             url = {https://elib.dlr.de/212711/}
}

@Article{Pflimlin2010a,
  author       = {Pflimlin, J.-M. and Binetti, P. and Souères, P. and Hamel, T. and Trouchet, D.},
  date         = {2010-03},
  year         = {2010},
  journal = {Control Engineering Practice},
  title        = {Modeling and attitude control analysis of a ducted-fan micro aerial vehicle},
  doi          = {10.1016/j.conengprac.2009.09.009},
  issn         = {0967-0661},
  number       = {3},
  pages        = {209--218},
  volume       = {18},
  publisher    = {Elsevier BV},
}

@Article{Pfister2024a,
  author       = {Pfister, Lena and Gohm, Alexander and Kossmann, Meinolf and Wieser, Andreas and Babić, Nevio and Handwerker, Jan and Wildmann, Norman and Vogelmann, Hannes and Baumann-Stanzer, Kathrin and Alexa, Almut and Lapo, Karl and Paunović, Ivan and Leinweber, Ronny and Sedlmeier, Katrin and Lehner, Manuela and Hieden, Alexander and Speidel, Johannes and Federer, Maria and Rotach, Mathias W.},
  date         = {2024-07},
  journal = {Meteorologische Zeitschrift},
  title        = {The TEAMx‑PC22 Alpine field campaign – Objectives, instrumentation, and observed phenomena},
  doi          = {10.1127/metz/2024/1214},
  issn         = {0941-2948},
  number       = {3},
  pages        = {199--228},
  volume       = {33},
  year         = {2024},
  publisher    = {Schweizerbart},
}

@Article{Segales2025,
  author    = {Segales, Antonio R. and Bell, Tyler M. and Tasim, Abdullah A. and Quiroz, Aaron and Simms, Jeremy D. and Gebauer, Joshua and Smith, Elizabeth N.},
  date      = {2025-12},
  title     = {CopterSonde-SWX: Development of a UAS-based Vertical Atmospheric Profiler for Severe Weather},
  doi       = {10.5194/egusphere-2025-4843},
  publisher = {Copernicus GmbH},
  year         = {2025},
}

\end{document}